\documentclass[12pt,letterpaper,english]{article}

 \usepackage{setspace,graphics}
 \usepackage[dvips]{epsfig} 
 \usepackage{a4,amssymb,epsfig,array,cite}
 \usepackage[hypertex]{hyperref}

\usepackage{amsfonts,bm,amssymb,euscript,array,babel}

\newcounter{multieqs}

\newcommand{\be}{\begin{equation}}
\newcommand{\ee}{\end{equation}}
\newcommand{\eq}[1]{(\ref{#1})}

\def\nn{\nonumber}
\def\bea{\begin{eqnarray}}
\def\eea{\end{eqnarray}}
\def\obar{\overline}

\def\beqa{\begin{eqnarray}} 
\def\eeqa{\end{eqnarray}} 
\def\beq{\begin{equation}} 
\def\eeq{\end{equation}}

\def\Tr{{\rm Tr}}

\def\a{\alpha}          
           
  \def\C{\Gamma}

\def\l{\lambda} \def\L{\Lambda}

\def\cA{{\cal A}}  \def\cC{{\cal C}}
  \def\cF{{\cal F}}
 \def\cH{{\cal H}} 
  
\def\cM{{\cal M}} \def\cN{{\cal N}} 
  
  \def\cU{{\cal U}}

\def\R{{\mathbb R}}
\def\C{{\mathbb C}}

\def\one{\mbox{1 \kern-.59em {\rm l}}}

\def\msu{\mathfrak{s}\mathfrak{u}}
\def\mmu{\mathfrak{u}}

\def\bit{\begin{itemize}}
\def\eit{\end{itemize}}

\def\({\left(}
\def\){\right)}

\def\uno{\mbox{1 \kern-.59em {\rm l}}}

\def\bcomment#1{}

\renewcommand{\title}[1]{\vspace{10mm}\noindent{\Large{\bf #1}}\vspace{8mm}}
\newcommand{\authors}[1]{\noindent{\large #1}\vspace{5mm}}
\newcommand{\address}[1]{{\itshape #1\vspace{2mm}}}

\begin{document}

\begin{titlepage}

\begin{flushright}
UWTHPh-2007-17\\
\end{flushright}

\begin{center}
  
\title{Emergent Gravity from Noncommutative Gauge Theory}

\authors{Harold {\sc Steinacker}${}^{1}$}

\address{ Fakult\"at f\"ur Physik, Universit\"at Wien\\
Boltzmanngasse 5, A-1090 Wien, Austria}

%\footnotetext[1]{harald.grosse@univie.ac.at}
\footnotetext[1]{harold.steinacker@univie.ac.at}

\vskip 2cm

\textbf{Abstract}

\vskip 3mm 

\begin{minipage}{14cm}%

We show that the matrix-model action 
for noncommutative $U(n)$ gauge theory
actually describes $SU(n)$ gauge theory coupled to 
gravity. This is elaborated in the 4-dimensional case.
The $SU(n)$ gauge fields as well as additional scalar fields 
couple to an effective metric $G_{ab}$, which is 
determined by a dynamical Poisson structure.
The emergent gravity is intimately related to noncommutativity,
encoding those degrees of freedom 
which are usually interpreted as $U(1)$ gauge fields. 
This leads to a class of metrics which contains the
physical degrees of freedom of gravitational waves, and allows to 
recover e.g. the Newtonian limit with arbitrary mass distribution.
It also suggests a consistent picture of UV/IR mixing in terms
of an induced gravity action. This should provide 
a suitable framework for quantizing gravity.

\end{minipage}

\end{center}

\end{titlepage}

\setcounter{page}0
\thispagestyle{empty}
%\newpage

%%%%%%%% table of content %%%%%%%
\begin{spacing}{.3}
{
\noindent\rule\textwidth{.1pt}            % THEN MAKE TOC...
   \tableofcontents
\vspace{.6cm}
\noindent\rule\textwidth{.1pt}
}
\end{spacing}

%%%%%%%ordinary document (end) ####################################

\section{Introduction}

It is generally accepted that the classical concepts of space and time 
will break down at the Planck scale, where quantum fluctuations of
space-time due to the interplay between gravity and quantum mechanics
become important. 
One way to approach this problem is to 
replace classical space-time by some kind of quantum space,
incorporating space-time uncertainty relations such as those 
obtained in \cite{Doplicher:1994tu}. This leads to  
noncommutative (NC) field theory, where some fixed NC space is
assumed; for basic reviews see e.g. \cite{reviews}. 

After considerable progress in the understanding of field
theory on  ``fixed`` NC or quantum spaces,
it is of fundamental importance to understand how a dynamical
quantum space in the
spirit of general relativity can be incorporated in such a framework. 
If noncommutative spaces are related to quantum gravity,
the incorporation of gravity should be simple and natural. 
Furthermore, one should take into account the lessons from string
theory, 
which provides a realization of quantum spaces as D-branes in
a nontrivial $B$-field background \cite{Seiberg:1999vs}, 
and points to a relation with gravity 
\cite{Kabat:1997sa,Banks:1996vh,Ishibashi:1996xs,Bigatti:1997jy,
Aoki:1999vr,Ishibashi:2000hh,Kimura:2000ur,Kitazawa:2006pj}. 
While several formulations of NC gravity have been proposed 
by deforming various formulations of general relativity 
\cite{Aschieri:2005zs,Nair:2006qg,Calmet:2005qm,Madore:2000aq,
Cardella:2002pb,Langmann:2001yr,Szabo:2006wx,Kurkcuoglu:2006iw,Mukherjee:2006nd}, 
a simple and compelling mechanism would be very desirable.

To identify this mechanism, it is helpful to reconsider 
gauge theories. There is a very simple and natural formulation of 
$\mmu(n)$ NC gauge theory in terms of matrix models, 
typically of the form
$S \cong Tr [X^a,X^b][X^a,X^b] + ... $ . 
Such actions describes gauge theory
on the quantum plane $\R^d_\theta$.  Similar actions arise
in the context of string
theory, such as the IKKT model \cite{Ishibashi:1996xs}.
The dynamical objects are  matrices resp. operators 
$X^a= Y^a + A^a\,\in\,\cA \otimes \mmu(n)$ 
("covariant coordinates"), 
where $Y^a$ generates the algebra of functions
$\cA\cong L(\cH)$ on  some NC  space.
The central observation is that the fluctuations $A^a \in \cA$
of the covariant coordinates can be interpreted
as $\mmu(n)$-valued gauge fields on the NC space.  
These considerations become more rigorous 
for compact quantum spaces 
such as $\C P_N^2$ or $S^2_N \times S^2_N$,
which are described by {\em finite} matrix models 
of similar type \cite{Behr:2005wp,Kitazawa:2002xj}. 

Even though this realization of gauge fields is very appealing, 
it is nevertheless strange: 
fluctuations of NC coordinates ought to describe 
fluctuations of the geometry, rather than  gauge fields.
This is particularly compelling for gauge 
theory ``on'' fuzzy spaces 
such as $\C P_N^2$ or $S^2_N \times S^2_N$, 
where the geometry of the space is indeed
dynamical and given by the minimum $\langle X^a \rangle = Y^a$ 
of an appropriate  
matrix model. This strongly hints at an implicit
gravity sector. There is also strong evidence for the presence of 
gravity in the IKKT matrix model of type IIB string theory
\cite{Ishibashi:1996xs,Aoki:1999vr,Ishibashi:2000hh,Kimura:2000ur,
Kitazawa:2006pj}, 
and even for a D=4 compactification thereof 
\cite{Ishibashi:2000hh,Kitazawa:2006pj} 
which can be viewed as a supersymmetric version 
of the model which will be studied here.
Further striking parallels between gravity and NC gauge theory
include the absence of local observables, and the implementation of
translations as gauge transformations. 
Finally, the $\mmu(1)$ sector of D=4 noncommutative gauge theory
is afflicted by the infamous UV/IR mixing 
\cite{Minwalla:1999px,Matusis:2000jf,Hayakawa:1999zf}, 
leading to a behavior which is very different from electrodynamics.

We show in this paper that the matrix model formulation of 
NC gauge theory in 4 dimensions does in fact contain gravity. 
More precisely, it 
should be interpreted as $\msu(n)$ gauge theory coupled to
gravity, with dynamical geometry  determined by
$\mmu(1)$ components of the covariant coordinates $X^a$.  This
solves at the same time a long-standing problem how to define 
NC $\msu(n)$ gauge theory: It has been known that the $\mmu(1)$ sector of
NC gauge theory cannot be disentangled from the $\msu(n)$ sector in any
obvious way. Here we understand this fact as the coupling of the
$\msu(n)$ gauge fields to gravity. 

One may wonder how it is possible that
nontrivial geometries arise from what is usually interpreted as 
$\mmu(1)$ gauge fields.
The answer is  quite simple: the effective geometry is
determined by the metric 
$G^{ab} = -\theta^{ac}(y) \theta^{bd}(y) g_{cd}$,
where $\theta^{ac}(y)= \obar\theta^{ac} + F^{ab}(y)$ 
is the dynamical Poisson tensor which is usually split into
background NC space and $\mmu(1)$ field strength, 
for $g_{ab} = \delta_{ab}$ 
resp. $g_{ab} = \eta_{ab}$ in the Euclidean resp. Minkowski case.
While such metrics do not reproduce the most general geometries, they
do contain the physical degrees of freedom of gravitational waves, 
and allow to obtain e.g. the Newtonian limit.
Therefore this provide a physically viable class of geometries
for gravity.

The observation that gravity can arises from NC gauge theory is not new. 
In particular, Rivelles \cite{Rivelles:2002ez} found a linearized version of 
the same effective metric 
coupling to scalar fields on $\R^4_\theta$, 
without addressing however nonabelian gauge fields. 
The idea that NC $\mmu(1)$ gauge theory should be viewed as gravity 
was put forward explicitly in
\cite{Yang:2004vd,Banerjee:2004rs} from the string theory point of view. 
We establish this mechanism in detail  
based on a very simple and explicit matrix model, and clarify the
associated geometry. 

The basic message is that gravity is already contained in the simplest
matrix models of NC gauge theory. There is no need to invoke
any new ideas.
This striking mechanism takes advantage of noncommutativity in 
an essential way, and has no commutative analog. 
Furthermore, the quantization of matrix models is naturally defined
by integrating over the space of matrices.
We will argue that this induces the 
action for gravity in the spirit of 
\cite{Sakharov:1967pk}, which suggests a 
natural role of UV/IR mixing.
However, the vacuum equations $R_{ab}\sim 0$ are 
obtained even at tree level.
While some freedom remains for modification of the
action (in particular extra dimensions),
the resulting gravity theory appears to be quite rigid. 
It is different from
general relativity but consistent with the Newtonian limit.
Moreover, some post-Newtonian corrections of general relativity 
 appear to be reproduced,
however a more detailed analysis is required.
While no final judgment can be made here concerning 
the physics, 
simplicity and naturalness certainly support this mechanism. 

The results of this paper should also shed
new light on  gravity in the IKKT model, 
 in the presence of a noncommutative D-brane. While 
this model is expected to contain gravity due to its
relation with string theory, an 
explicit identification of nontrivial geometries has proved 
to be difficult \cite{Hanada:2005vr,Kawai:2007tn}. This is discussed
in section \ref{sec:metric}.

The outline of this paper is as follows. We first explain the
separation of the covariant coordinates in geometric and gauge 
degrees of freedom, which is the essential step of our approach.
This leads to a dynamical theory of Poisson manifolds, 
to which we associate in section \ref{sec:metric} an effective
metric. In section \ref{sec:SW} we establish
that this metric indeed governs the low-energy behavior
of both scalar and gauge fields. The technical details for the gauge fields
are lengthy and delegated to Appendix A.
Section \ref{sec:induced-grav} elaborates to some extent 
the physical content of the 
emerging gravity theory, in particular the induced
Einstein-Hilbert-like action, UV/IR mixing, gravitational waves, 
the Newtonian limit and few examples. We conclude with discussion and outlook.

\section{Gauge fields and Poisson geometry}

Consider the following matrix model 
action for noncommutative gauge theory in 4 dimensions
\be
S_{YM} = - Tr [X^a,X^b] [X^{a'},X^{b'}] g_{a a'} g_{b b'}, 
\label{YM-action-1}
\ee
for
\be
g_{a a'} = \delta_{a a'} \quad \mbox{or}\quad g_{a a'} = \eta_{a a'} 
\label{background-metric}
\ee
in the Euclidean  resp.  Minkowski case.
Here the "covariant coordinates"  $X^a$ are hermitian matrices 
or operators acting on some Hilbert space $\cH$.
The basic symmetries of this action are the gauge symmetry
\be
X^a \to U X^a U^{-1}, \qquad U \, \in \,\cU(\cH)
\label{NCgaugetrafo}
\ee
where $\cU(\cH)$ are the unitary operators on $\cH$, 
translational invariance
$X^a \to X^a + c^a$
for $c^a \in \R$, and global $SO(4)$ resp. $SO(3,1)$ invariance.
The more conventional action 
$Tr([X^a,X^b]-\obar\theta^{ab})^2$ 
for $\R^4_{\obar\theta}$ \cite{Madore:2000en} differs from
\eq{YM-action-1} only by a constant shift and 
a topological or boundary term  of the form
$Tr[X^a,X^b]\, \obar\theta^{ab}$.
We consider \eq{YM-action-1} to avoid introducing 
the constant tensor $\obar\theta^{ab}$ at this point,
thereby stressing background-independence. 
This is also the type of action which is typically 
found in the context of string theory 
\cite{Ishibashi:1996xs,Seiberg:1999vs}.
The equations of motion are 
\be
[X^a,[X^{a'},X^{b'}]] g_{a a'} =0 .
\label{eom-0}
\ee
A particular solution is given by $X^a = \obar Y^a$,
where the $\obar Y^a$  satisfy the commutation relations
\be
[\obar Y^a,\obar Y^b] =  \obar\theta^{ab} .
\label{moyal-weyl}
\ee
These generate the algebra $\cA \cong \R^4_{\obar\theta}$
of functions on the Moyal-Weyl quantum plane.
Here $\obar\theta^{ab}$ is assumed to be
constant and non-degenerate, and 
the $\obar Y^a$ have the standard Hilbert-space representations.
To avoid cluttering the formulas with $i$  we adopt the 
convention that
$\theta^{ab}$ is purely imaginary, and similarly for 
the field strength $F_{ab}$ etc. below.
Another solution is given by $X^a = \obar Y^a \otimes \one_n$,
which will lead to $\mmu(n)$ gauge theory\footnote{the rank $n$ of 
therefore not determined by the matrix model but by the 
choice of vacuum solution. If desired, $n$ can 
be controlled at least in the Euclidean case by compactifying the space
and considering e.g. fuzzy $S^2 \times S^2$ or $\C P^2$ 
\cite{Behr:2005wp}, where $\cH$ is finite-dimensional.}.

In this paper, we will focus on configurations (which need not be
solutions of the e.o.m.)
which are close to the ``vacuum'' solution  
$X^a = \obar Y^a \otimes \one_n$. This will lead to  
noncommutative $\mmu(n)$ gauge theory, or rather 
$\msu(n)$ gauge theory coupled to gravity. 
%We stress that the algebra $\cA$ itself will {\em not} determine the 
%geometry, and its details are not important here. 
%Therefore we consider for simplicity
Hence consider small fluctuations of the form
\be
X^a = \obar Y^a\otimes \one_n + \cA^a(\obar Y)
\ee
with $\cA^a(\obar Y) \in \cA\otimes M_n(\C)$.
We will replace $f(\obar Y) \to f(\obar y)$ whenever
$f(\obar Y) \in \cA$ can be well approximated by a classical function 
$f(\obar y)$. In the conventional interpretation, 
$\cA^a(\obar Y) = \cA^{a}_0(\obar Y)\otimes \one_n +
\cA^a_\a(\obar Y)\otimes \tau_\a$ 
is viewed as $\mmu(n)$-valued
gauge field, where $\tau_\a$ are a basis of $\msu(n)$.
Here we will adopt a different approach, 
separating the trace- $\mmu(1)$ part 
(i.e. the coefficient of $\one_n$)
and the remaining nonabelian part as follows:
\be
X^a  =\,\, Y^{a} \,\one_n \,\, +  \,\, \cA^{a}(Y)\,\,
     =  Y^{a} \,\one_n \,\, +  \,\, \cA^{a}_\a(Y) \,\tau_\a\,\, .
\label{gaugefield-splitting}
\ee
Here
\be
Y^{a} = \obar Y^a + \cA^{a}_0(\obar Y)
\ee
contains the full trace-$\mmu(1)$ component,
and will be interpreted as generators of a NC space 
$\cM_\theta$ with general noncommutativity 
\be
[Y^a,Y^b]\,\, \equiv\,\, \theta^{ab}(Y)\,\,\approx\theta^{ab}(y) \, .
\label{theta-general}
\ee  
The other, nonabelian components $\cA^{a}_\a(Y)$ 
will be considered as functions of the coordinate generators $Y^a$
resp. $y^a$.

The essential point is the following: what is usually interpreted as  
``abelian gauge  field'' $\cA^{a}_0$ is understood here as
fluctuation of the quantum space, which 
determines a Poisson structure $\theta^{ab}(y) $ and eventually
a metric $G^{ab}(y)$ \eq{effective-metric}. The remaining ``nonabelian'' 
$\cA^{a}_\a(y) \,\tau_\a$  describe $\msu(n)$-valued gauge field. 
The well-known fact that the $\mmu(1)$ and $\msu(n)$ components
cannot be completely disentangled in NC gauge theory
will be understood here as coupling of the $\msu(n)$ gauge fields
to gravity.

 The physical reason why the splitting \eq{gaugefield-splitting} 
of $\mmu(1)$ and $\msu(n)$ components is appropriate 
 will be seen by considering 
gauge-invariant actions such as  \eq{YM-action-1} or
\eq{scalar-action-0}.
The reason is that the kinetic term in the underlying
matrix-model action always involves the
induced metric $G^{ab}$ identified below. 
This universal coupling to a metric $G^{ab}$ is strongly
suggestive of gravity. 
This is based on the observation that in the framework of matrix
models, all fields must be in the adjoint in order to acquire a kinetic term.
However, other types of matter and 
low-energy gauge fields close to those required for the standard model
can arise after spontaneous symmetry breaking, 
see e.g. \cite{Aschieri:2006uw}.

\paragraph{Semi-classical limit: Poisson manifolds.}

We want to understand the geometrical significance of 
the various configurations \eq{theta-general}. 
The emerging picture is that the
$\mmu(1)$ sector of the matrix model 
describes a dynamical theory of Poisson manifolds.

Consider 
generators $Y^a$ of $\cA$ satisfying  \eq{theta-general}, 
and assume that $\theta^{ab}(Y)$ is ``close'' to
a smooth Poisson structure $\theta^{ab}(y)$.
This defines a (local) Poisson manifold $(\cM,\theta^{ab}(y))$
whose quantization is given by $Y^a$.
Conversely, using a general result of Kontsevich \cite{Kontsevich:1997vb}
we can quantize essentially any Poisson structure\footnote{we 
ignore the distinction between formal and convergent star products here.}
at least locally via such $Y^a$.
To make this mathematically more precise, the 
concept of a star-product is useful.  Given 
an isomorphism of vector spaces
\be
\cC(\cM) \to \cA \, 
\ee
where $\cC(\cM)$ denotes the space of functions on $\cM$,
one can define via pull-back a ``star product'' on  $\cC(\cM)$. 
Assuming that this star product has a meaningful
expansion in powers of $\theta$, the
commutator of 2 elements in $\cA$ reduces to 
the Poisson bracket of the classical functions on $\cM$ to leading
order in $\theta$. More precisely,
using a suitable change of variables
one can choose the star product (e.g. by taking the one given in
 \cite{Kontsevich:1997vb}) such that 
\be
[f,g] = i\{f,g\} \,\,+ O(\theta^3)  \,\,=  \,\,
\theta^{ab}(y)\,\partial_a (f)\, \partial_b (g)  \,\,+ O(\theta^3)
\ee
to $O(\theta^3)$. This will be important below 
in order to extract the semiclassical limit.
In particular, this implies
\be
 [Y^a, f(Y)] =  i\{y^a, f(y)\}  \,\, + O(\theta^3) 
\,\, =\, \theta^{ab}(y)\, \partial_b f(y)  \,\, + O(\theta^3) \, .
 \label{deriv-inner}
\ee
where $y^a$ denotes the pull-back of $Y^a$.

\section{Effective Metric}
\label{sec:metric}

We now show how a dynamical metric arises naturally from matrix model
actions. The basic mechanism is seen most easily for scalar fields.

\paragraph{Scalar fields}

In the framework of matrix models, the only possibility to obtain
kinetic terms is through commutators 
$[X^a,\Phi] \sim \theta^{ab}(y) \partial_b \Phi + [\cA^a,\Phi]$. Therefore only
 fields in the adjoint are admissible, with action
\be
S[\Phi] = - Tr\, g_{aa'} [X^a,\Phi][X^{a'},\Phi] .
\label{scalar-action-0}
\ee
In a configuration as in  \eq{gaugefield-splitting}
with nontrivial background $Y^a$ 
and $\msu(n)$-valued fluctuations 
\be
X^a = Y^{a}\otimes \one_n + \cA^a(Y)\, ,
\ee
we can obtain the commutative limit 
using the naive change of variables 
\be
\cA^a = \theta^{ab}(y) \tilde A_b
\label{A-naive}
\ee
where $\tilde A_a$ is antihermitian. The action then takes the form
\bea
S[\Phi] &\approx& - Tr\, \theta^{ab}(y)\,\theta^{a'c}(y) g_{aa'}\,
 (\partial_b\Phi + [\tilde A_b,\Phi]) 
(\partial_c \Phi + [\tilde A_c,\Phi])\nn\\
&=&  Tr\, G^{ab}(y)\,  D_a\Phi D_b \Phi
\label{scalar-action}
\eea
to leading order, defining the effective metric
\be
G^{ab}(y) = -\theta^{ac}(y) \theta^{b d}(y)\, g_{cd}
\label{effective-metric}
\ee
where $g_{cd}$ is the background metric \eq{background-metric}
and $D_a = \partial_a + [\tilde A_a,.]$.

Some remarks are in order. We will show below that the naive
substitution \eq{A-naive} is  sufficient here and \eq{scalar-action} is
indeed the correct classical limit. 
An infinitesimal version of \eq{effective-metric} 
was already obtained in \cite{Rivelles:2002ez} up to 
a trace contribution, which is explained in section \ref{sec:flat-expansion}.
Furthermore, observe that
\be
e^a_b(y) := -i\theta^{ac}(y) g_{cb}
\label{vielbein}
\ee
can be interpreted as vielbein; 
this is consistent with the expression \eq{effective-metric} 
for the metric $G^{ab}$.
The antisymmetry of $\theta^{ac}(y)$ reflects the choice of a special
``gauge'' in comparison with the standard formulation of 
general relativity.
Note that $G^{ab}$ is nondegenerate if and only if the
Poisson tensor $\theta^{ab}(y)$ is non-degenerate. In this paper we
assume that $\theta^{ab}(y)$ is non-degenerate, even though degenerate
cases are possible and are expected to be very interesting.
Finally, the effective metric $G^{ab}$ determines in particular 
the spectrum of the Laplacian acting on $\Phi$;
this will become important in section \ref{sec:induced-grav}.

\paragraph{Nonabelian gauge fields.}

Now consider the commutator $[X^a,X^b]$ 
in the nonabelian case, for the same background. 
Using \eq{gaugefield-splitting}, we have
\be
[X^a,X^b] = \theta^{ab}(Y)\,\one_n + \cF^{ab}(Y) 
\label{commutator-split}
\ee
where 
\bea
\cF^{ab} &=& [Y^a,\cA^b] - [Y^b,\cA^a] + [\cA^a,\cA^b] 
\eea
is the noncommutative field strength. 
Our aim is to obtain the classical limit of the action
\eq{YM-action-1}, and to show that it can be interpreted
as an ordinary gauge field coupled to the effective metric
$G^{ab}$. To develop some intuition, we first 
give a naive, incomplete argument before embarking into the 
correct but less transparent Seiberg-Witten expansion.

\paragraph{Naive analysis}

Let us try the naive\footnote{replacing this with the slightly less-naive 
$\cA^a =\frac 12 \{\theta^{ab}(y), \tilde A_a \}$ does not 
solve the problem} change of variables \eq{A-naive}
which for constant $\theta^{ab}$ correctly leads to the
classical limit.
Using \eq{deriv-inner}, we would obtain
\bea
\cF^{ab} &=& \theta^{ac}(y)\theta^{bd}(y)\, \cF_{cd}
\eea
where
\be
\cF_{ab} \, \approx\, \partial_a  \tilde A_b(y)
 - \partial_b \tilde A_a(y) + [\tilde A_a(y), \tilde A_b(y)]
\quad + O(\theta^{-1} \partial \theta) \,\, + O(\theta)
\label{F-naive}
\ee
where $O(\theta^{-1} \partial \theta)$ stands for terms of the type
$\theta^{-1}_{cd}[\theta^{db},\tilde A_a]$. These are small
as long as
\be
\theta^{-1}\, \partial \theta \ll \partial \tilde A_a,
\ee
i.e. if the variations of $\theta^{ab}(y)$ resp. $G^{ab}$ are
much slower than those of the gauge fields $\tilde A_c$.
One can then interpret $\cF_{ab}(y)$
as gauge field strength, which certainly holds  for constant $\theta^{ab}$.
Note also that the leading term of $\cF_{ab}$ 
takes values in $\msu(n)$, but there are $\mmu(1)$
contributions of order $\theta$ due to e.g. $\{\cA^a,\cA^b\}$.
Neglecting these, we would have
$$
Tr (\theta^{ab} \cF^{ab}) \approx 0 
$$
and the action would be  
\bea
S_{YM} &\approx & -Tr \(\theta^{ab}\, \theta^{a'b'} 
+ \cF^{ab}(y)\cF^{a'b'}(y)\) g_{aa'} g_{bb'} \nn\\
&=& Tr  \(G^{ab}(y) g_{ab} - G^{c c'}(y)\, G^{d d'}(y)
 \, \cF_{cd}(y)\, \cF_{c'd'}(y) \)
\eea
in the semi-classical limit. This suggests that
the nonabelian gauge fields are indeed coupled as expected to the
open-string metric $G^{ab}$. However, we need a more sophisticated 
analysis using the Seiberg-Witten map to establish this, because
the neglected terms in \eq{F-naive}
are of the same order as the coupling to the gravitational 
fields i.e. the connection, and the $\mmu(1)$ 
terms in $\{\cA^a,\cA^b\}$ do in fact contribute at the leading order.
This is reflected by the fact that $\cF_{ab}$
is not gauge invariant in the commutative limit unless 
$\theta^{ab} = const$.

\paragraph{Relation with string theory.}

Our effective metric $G^{ab}$ \eq{effective-metric}
is strongly reminiscent of the ``open string metric'' on 
noncommutative D-branes in a $B$-field background  \cite{Seiberg:1999vs}, 
in the Seiberg-Witten decoupling limit $\alpha' \to 0$. 
Our background metric $g_{ab}$ 
can then be interpreted as ``closed string metric''  of the embedding
space. However, the $\theta^{ab}(y)$ which enters our metric
$G^{ab}$ is non-constant and determined by the full $\mmu(1)$ part of 
$B' = B+\cF$ on the brane, unlike in  \cite{Seiberg:1999vs}. 
This should be related to the symmetry
$A \to A+\Lambda, B \to B - d\Lambda$ in the context of string theory
as pointed out in \cite{Yang:2004vd,Banerjee:2004rs}, where the
different role of the $\mmu(1)$ and the $\msu(n)$ sectors was ignored however.
We will see below that $G^{ab}$ 
is also the effective metric for the $\msu(n)$ YM-action.

\subsection{Effective gauge theory and Seiberg-Witten map}
\label{sec:SW}

In this section, we implement the separation \eq{gaugefield-splitting}
of the $X^a$ in NC background $Y^a$ and $\msu(n)$ gauge fields, 
and carefully determine the classical limit of the
action \eq{YM-action-1}. The $\msu(n)$-valued components 
of $\cA^a$ will be expressed using a Seiberg-Witten map
in terms of classical $\msu(n)$-valued gauge fields $A_{a}$,
on a noncommutative background $\theta^{ab}(y)$ determined by
the $\mmu(1)$ components $Y^a$. The latter
eat up the ``would-be $\mmu(1)$ gauge fields'' and determine the metric $G^{ab}(y)$. 
Thus the full $\mmu(1)$ sector
determines the dynamical NC parameter 
$\theta^{ab}(y)$ and the geometry  $G^{ab}$, 
while the nonabelian $\msu(n)$ fields are expanded to leading
order in $\theta^{ab}(y)$
and  couple to $G^{ab}$. This analysis is surprisingly 
involved.

Let us rewrite the nonabelian gauge fields 
$\cA^a_\a = \cA^a_\a(A_{a})$ in terms of 
classical antihermitian $\msu(n)$-valued
gauge fields $A_{a}$ using the Seiberg-Witten 
map\footnote{The Seiberg-Witten map is used
simply as a change of the field coordinates. It
does not imply that we work in the framework of star-products. 
The non-hermitean version is used here for brevity, which is easily 
made hermitian.}  
\cite{Seiberg:1999vs}, dropping the index $\a$
from now on.
The classical gauge fields transform under $\msu(n)$ gauge
transformations as
\be
\delta_{cl} A_a = -i \partial_a \l + i[\l,A_a]
 = -i\partial_a \l + i[\l,\tau^\a]\,A_{a,\a} \, .
\ee
The appropriate SW-map
for general $\theta^{ab}(y)$ is given by \cite{Jurco:2000fb}
\bea
\cA^a &=&  \theta^{ab} A_b 
- \frac 12 (A_c [Y^c,\theta^{ad} A_d] +  A_c F^{ca}) \quad 
+O(\theta^3) \nn\\
&=:&  \theta^{ab} A_b + A_{SW,2}^a  \quad + O(\theta^3)
\label{SW-gaugefields}
\eea
and satisfies
\be
\delta_\L (X^a) =  i[\Lambda,Y^{a} + \cA^a] =  \delta_{cl} \cA^a
\ee
with the NC gauge parameter
\be
\Lambda = \l + \frac 12 \theta^{ab} (\partial_a \l ) A_b \, .
\ee
This means that the action \eq{YM-action-1} expressed in terms of
$A_a$ is invariant under the 
classical $\msu(n)$ gauge transformations acting on $A_a$.
This in turn implies that 
the action can be written as a function of the ordinary $\msu(n)$ field strength 
\bea
F^{ab} &:=& \theta^{ac} \theta^{bd} F_{cd} \,
= \theta^{bd}[Y^a, A_d] -\theta^{ac} [Y^b, A_c] 
+ \theta^{ac} \theta^{bd}[A_c,A_d] \,\, +  O(\theta^4) \, \nn\\
F_{ab} &:=& \partial_a A_b - \partial_b A_a + [A_a,A_b]
\label{F-explicit} 
\eea
In this section,
we adopt the convention that indices are raised and lowered with 
$\theta^{ab}$ rather than a metric, e.g. $A^a := \theta^{ab} A_b$ etc.
Note that it is $F_{ab}$  rather than $\cF^{ab}$ 
 which has the
correct classical limit as a 2-form for general $\theta^{ab}(y)$, and
the classical limit can only be understood correctly 
in terms of $F_{ab}$. The reader not interested in technical details
can jump to the resulting action \eq{S-YM-effective}.

\paragraph{Contribution to the action.}

We want to obtain the classical limit of the action \eq{YM-action-1}
$$
S = - Tr (\cF^{ab} \cF^{ab} + 2 \theta^{ab} \cF^{ab} 
 + \theta^{ab}\theta^{ab})  
$$
in terms of the $A_a$ or $F_{ab}$. This
requires keeping all terms of order 
$O(\theta^4)$. The NC field strength is 
\bea
\cF^{ab} &=& [Y^a,\cA^b] - [Y^b,\cA^a]  +  [\cA^a, \cA^b] \nn\\
&=&  [Y^a,A^b]
 -  [Y^b,A^a]+  [A^a, A^b]  +\cF^{ab}_{SW,2} +  O(\theta^4) 
\eea
where
\bea
\cF^{ab}_{SW,2} &=& [Y^a + A^a, A_{SW,2}^b] 
+ [A^a_{SW,2}, Y^b+A^b] + [A^a_{SW,2}A_{SW,2}^b]  \nn\\
&=& [X^a, A_{SW,2}^b] + [A^a_{SW,2}, X^b]
\quad +  O(\theta^4),
\eea
since $\cA^a_{SW,2} = O(\theta^2)$. 
We must carefully keep track of the $\mmu(1)$ components of 
$\cF^{ab}$ to order $\theta^3$ and the $\msu(n)$ components 
to order $\theta^2$. Dropping higher-order terms, one has
\bea
\cF^{ab} &=& [Y^a,A_d \theta^{bd}] - [Y^b,A_d \theta^{ad}] 
  + [A^a,A^b] + \cF^{ab}_{SW,2} \nn\\
&=& F^{ab} - A_c[Y^c,\theta^{ab}] 
 + [A^a,A^b] - \theta^{aa'}\theta^{be'}[A_{a'} ,A_{e'}] + \cF^{ab}_{SW,2}
\label{F-identity-one}
\eea
using the Jacobi identity for $\theta^{ab}$, and thus
\bea
S &=& -Tr \Big(F^{ab}F^{ab} - 2 F^{ab} A_c [Y^c,\theta^{ab}]
+ A_c [Y^c,\theta^{ab}] [Y^d,\theta^{ab}] A_d \nn\\
&& + 2 (F^{ab} - A_c[Y^c,\theta^{ab}])
 ([A^a,A^b] - \theta^{aa'}\theta^{be'}[A_{a'} ,A_{e'}]) \nn\\
&& + ([A^a,A^b] - \theta^{aa'}\theta^{be'}[A_{a'} ,A_{e'}])^2 
 + 2 \theta^{ab}[A^a,A^b]\Big) + S_{SW,2}
\label{action-exxpanded}
\eea
up to $O(\theta^4)$, dropping the constant 
$Tr \theta^{ab}\theta^{ab}$ for now. 
Here 
\be
S_{SW,2} \,\, = -\, Tr \,( 2\theta^{ab}\cF^{ab}_{SW,2} ) \,\,  
= \, - Tr\, (4 \theta^{ab} [X^a, A^b_{SW,2}] )\, ,
\label{SW-action}
\ee
and we routinely drop subleading terms and 
use identities such as 
$Tr\theta^{ab} [Y^a,A^b] =0$ since $A_a \in \msu(n)$.
Similarly, we can set $[A,\theta] = 0$ 
in the $O(A^3)$ and $O(A^4)$ terms to leading order. 
For example, 
\bea
[A^a,A^b] = \theta^{aa'}\theta^{bb'}[A_{a'},A_{b'}] \quad + O(\theta^3)
\label{AA-drop}
\eea
which simplifies \eq{action-exxpanded}. Note also that
the only contribution from $\theta^{ab} \cF^{ab}$
is the NC (Poisson bracket) contribution in $\theta^{ab} [A^a,A^b]$. 
Therefore
\bea
S = -Tr \Big(F^{ab}F^{ab} - 2 F^{ab} A_c [Y^c,\theta^{ab}]
+ A_c [Y^c,\theta^{ab}] [Y^d,\theta^{ab}] A_d
 + 2 \theta^{ab}[A^a,A^b]\Big) + S_{SW,2}  \, . % \nn\\
\label{action-expanded}
\eea
After a tedious computation (see Appendix A)
using  elementary trace-manipulations, one obtains
\bea
S &=& -Tr \Big(F^{ab} F^{ab} 
 - \theta^{ab}  F^{ab}\theta^{cd} F_{cd}
-  2 \theta^{ab}F^{ad}\theta^{-1}_{dc} F^{bc} 
 + \frac 18\theta^{ab}\theta^{ab}\theta^{cd}\theta^{ij}
\big(F_{cd}F_{ij} + 2 F_{id} F_{jc} \big)  \Big) \nn\\
&=& -Tr \Big( G^{c c'} G^{d d'} F_{cd}\, F_{c'd'}
 +  F_{a'b'} F_{cd} (\theta^{a'a} \theta^{ab}\theta^{bb'}) \theta^{cd} 
+ 2 F_{a'c} F_{b'c'}(\theta^{a'a}\theta^{ab}\theta^{bb'})\theta^{c'c}\nn\\  
&& + \frac 18\theta^{ab}\theta^{ab}\theta^{cd}\theta^{ij}
\big(F_{cd}F_{ij} + 2 F_{id} F_{jc} \big)  \Big) \nn\\
&\equiv&  -Tr \, G^{c c'} G^{d d'} F_{cd}\, F_{c'd'} + S_{NC}
\label{action-expanded-2}
\eea
which is exact to order $O(\theta^4)$.
This action is manifestly gauge invariant, and for $\theta^{ab} = const$
it reduces to the standard YM action $S=Tr F^{ab} F^{ab}$
up to boundary terms, as it should.
From now on, we no longer raise or lower indices with $\theta^{ab}$.

The ``noncommutative'' terms $S_{NC}$ can be 
simplified further by considering the following dual evaluation of 
the 4-form resp. totally antisymmetric  4-tensor
$\frac 12 (F \wedge F)_{ijkl} = (F_{ij} F_{kl} - F_{il} F_{kj} - F_{lj} F_{ki})$:
\be
\frac 12 (F\wedge F)_{ijkl} \tilde\theta^{ij}\theta^{kl}
= (F_{ij}\tilde\theta^{ij}) (F_{kl}\theta^{kl})
 +  2 F_{il} F_{jk} \tilde\theta^{ij}\theta^{kl} .
\label{FF-theta}
\ee
We note that for $\tilde \theta^{ij} 
= \theta^{ik}\theta^{kl}\theta^{lj}= (\theta g \theta g \theta)^{ij} $
these are precisely the terms in $S_{NC}$, and conclude
\be
 S_{NC} =
-Tr \frac 12 (F\wedge F)_{ijkl}\(\tilde \theta^{ij}\theta^{kl}
 + \frac 18 (\theta^{ab} \theta^{ab}) \theta^{ij} \theta^{kl} \) 
\ee
where 
$\theta^{ab} \theta^{ab} \equiv - G^{ab} g_{ab}$
upon reinserting $g$.
Since $F \wedge F$ is a 4-form, it only couples to the 
totally antisymmetrized components $(\tilde\theta\wedge\theta)^{ijkl}$
of $\tilde\theta^{ij}\theta^{kl}$,
which can be interpreted as dual 4-form. Because the space of
4-forms is one-dimensional, we must have 
$\tilde\theta\wedge\theta = \eta(y) \theta\wedge\theta$, and it is
easy to see that (see Appendix C) 
\be
\eta(y) = \frac 14 \,G^{ab} g_{ab} \,.
\label{eta-eval}
\ee
Using  $(F\wedge F)_{ijkl}\theta^{ij}\theta^{kl} 
= \frac 16 (F\wedge F)_{ijkl} (\theta\wedge\theta)^{ijkl} 
= -\frac 13 \sqrt{\det(\theta^{ab})}(F\wedge F)_{ijkl} \,\varepsilon^{ijkl}$ 
we finally obtain
\be
 S_{NC} =
\frac 12  Tr (G^{ab} g_{ab})  \sqrt{\det(\theta^{ab})}\,
 \frac 1{4!} (F\wedge F)_{ijkl}\varepsilon^{ijkl} \, .
\ee
This reduces  to a topological 
surface term for constant $\theta^{ab}$, but not for general $\theta^{ab}(y)$.

\paragraph{Volume element.}

Finally we want to rewrite the trace as an integral in the semiclassical
limit. According to standard Bohr-Sommerfeld quantization, the 
appropriate relation should be 
\be
(2\pi)^2 Tr f(y) \sim \int \frac 12 \omega^{2} f(y)
= \int d^4y\, \rho(y)\, f(y)
\label{volume-density}
\ee
or equivalently
\be
(2\pi)^2 Tr \sqrt{\det(\theta^{ab})} \sim \int d^4 y ,
\ee
where $\omega = i\theta^{-1}_{ab}(y) dy^a dy^b$ is the symplectic form,
and $\frac 12 \omega^{2} = \rho(y) d^4 y$ 
the symplectic volume element.
A precise way to justify this for general (non-degenerate) 
$\theta^{ab}(y)$
is to require the trace property 
\be
\Tr [f,g] \sim \int\rho(y) \{f,g\} =0
\ee
up to boundary terms, which fixes $\rho(y)$ up to a constant factor.
It is easy to see that $ \rho(y) d^4 y  =  \frac 12\omega^2$ indeed
satisfies this requirement: 
\bea
\int \omega^2 \{f,g\} &=& \int \omega^2 X_f[g] 
= \int \omega^2 i_{X_f} dg \nn\\
&=& - \int (i_{X_f}\omega^2)  dg =2 \int (i_{X_f}\omega) \omega dg 
= \int df \omega dg =0
\eea
up to boundary terms,
where $X_f$ is the Poisson vector field generated by $\{f,.\}$.
Explicitly,
\bea
\rho(y) &=& \rm{Pfaff}(i\theta^{-1}_{ab}) = \sqrt{\det{\theta^{-1}_{ab}}}
= \(\det(g_{ab}) \det(G_{ab})\)^{1/4} \nn\\
&=:& \Lambda_{NC}^{4}(y) 
\label{rho}
\eea
where $\Lambda_{NC}(y)$ can be interpreted as ``local'' scale of noncommutativity.

\paragraph{Effective gauge action.}

Reinserting the constant term 
\be
-Tr\, \theta^{ab} \theta^{a'b'} g_{aa'} g_{bb'}
=  Tr\, G^{aa'} g_{aa'} = 4 Tr\, \eta(y)
\ee
we finally obtain the 
classical limit of the action \eq{YM-action-1} 
in the background $Y^a$:
\be\fbox{$
S_{YM} = c \int d^4 y\, \rho(y) 
tr \(4\eta(y) - G^{c c'} G^{d d'} F_{cd}\, F_{c'd'}\)
 + 2 c \int \eta(y) \, tr F\wedge F $}
\label{S-YM-effective}
\ee 
where an overall constant $c$ has been inserted, and $tr()$ denotes
the trace over the $\msu(n)$ components. 
This is an action for a $\msu(n)$ gauge field coupled to a 
dynamical metric
$G^{ab}(y)$ and the constant background metric $g_{ab}$.

Note that $S_{YM}$ 
is invariant under local Lorentz transformations, if we consider
$\eta(y)$ as a scalar functions.
This is remarkable, because it can be viewed as a re-summation of a 
Seiberg-Witten expansion in $\mmu(1)$ 
from the Moyal-plane point of view, 
where it would appear to suffer from Lorentz violation. Therefore
predictions and apparent problems for gauge theories 
on $\R^4_{\obar \theta}$
due to apparent Lorentz-violation may largely disappear here.

It is fascinating to observe that $\eta(y)$ takes the place of
both the cosmological ``constant'' and the axion, which is
related to the strong CP problem. To explain that both are small
are outstanding problems. The theory emerging here
is expected to have important consequences on these issues, 
however this can only be addressed after quantum effects are taken
into account. We will see that the first term  
in fact should {\em not} be interpreted as cosmological constant,
rather it leads to the vacuum equations of motion at tree level 
\eq{eom-tree}, \eq{ricci-flat}.
 Very similar actions have been considered from the
classical point of view in 
\cite{Alvarez:2005iy,Henneaux:1989zc}, however with an independent field
replacing $\eta(y)$.

We also note that \eq{rho} implies the relation
$(2\pi)^2 \cN = (2\pi)^2  Tr \one = \int d^4 y\, \Lambda^{4}_{NC}(y)$,
where $\cN$ is the dimension of the 
underlying Hilbert space $\cH$ in the compact case.  Therefore the local scale of
noncommutativity can be interpreted as ``local'' dimension of $\cH$ per
coordinate volume,
\be
\Lambda_{NC}^4 \sim \frac{(2\pi)^2\cN}{\rm{Vol}} .
\ee

\paragraph{Scalar field.}

Similarly, we want to obtain the 
classical limit of the scalar action \eq{scalar-action-0}.
Strictly speaking, we should also use a Seiberg-Witten map for the
scalars, in order to get the correct gauge-invariant classical 
limit. This is given by 
\be
\Phi = \phi - \theta^{ab} A_a \partial_b \phi 
- \frac 14\theta^{ab} [A_a A_b,\phi] + O(\theta^2) 
\ee
Noting that
\be
[X^a,\Phi] = [Y^a + \theta^{ab}A_b,\phi + O(\theta)]
= \theta^{ab}(\partial_b \phi + [A_b,\phi]) + O(\theta^2)
\ee
we obtain
\be \fbox{$
S[\Phi] %  &=& Tr G^{ab}(\partial_b \phi + [A_b,\phi]) 
% (\partial_b\phi + [A_b,\phi]) + O(\theta^3) \nn\\ 
=  \int d^4 y\, \rho(y)\, tr\, G^{ab} (\partial_b \phi + [A_b,\phi]) 
(\partial_b\phi + [A_b,\phi]) $}
\label{S-scalar-eff} 
\ee 
to leading order.
Therefore in the scalar case, the correct classical limit
is indeed obtained by the naive analysis 
leading to \eq{scalar-action}. 
In particular, we obtain the same effective metric $G^{ab}$
coupling to both scalar and gauge fields. 
This is of course essential for an interpretation 
in terms of gravity, and resolves an inconsistency for
the gauge fields in \cite{Rivelles:2002ez}. Note furthermore the
invariance of \eq{S-scalar-eff} under Weyl rescaling $G \to e^\sigma\,
G$, which is usually found for the Yang-Mills sector.

The effective actions \eq{S-YM-effective} and \eq{S-scalar-eff}
almost have the standard form of gauge resp. scalar fields
coupled to an external metric $G^{ab}$, except for the 
density functions $\rho(y)$ and $\eta(y)$ which depend not only 
on  $G_{ab}$ but also on the ``background'' or
closed string metric $g_{ab}$. If we consider $g_{ab}$ as a metric tensor,
then these actions are generally covariant.
However, $g_{ab}$ is a fixed matrix in the fundamental action \eq{YM-action-1},
where it does not make sense to transform it under a  general
diffeomorphisms. 
Thus general covariance arises only in the effective
low-energy action, considering $g_{ab}$ as a background metric
which enters the Yang-Mills action only through $\det g_{ab}$ and $\eta(y)$. 
For fixed $g_{ab}$, the
Yang-Mills term in \eq{S-YM-effective} is covariant only under
volume-preserving diffeomorphisms, and the 
``would-be topological'' correction term $S_{NC}$ is invariant under
diffeomorphisms preserving $\eta(y)$. 
This is somewhat reminiscent of
unimodular gravity \cite{Einstein:1919gv}, but more restrictive.

It may be tempting to recover the ``missing'' density factor in \eq{S-YM-effective}
by defining a slowly varying effective gauge coupling for the
Yang-Mills sector, 
\be
\frac 1{G^2_{\rm YM}(y)} = c \, \Big(\frac{\det g_{ab}}{\det G_{ab}}\Big)^{1/4}\, .
\ee
However this is premature and perhaps misleading at this point, 
because a similar $\msu(n)$ action will be induced at one-loop,
which might have a different density factor.

\section{Emergent Gravity}
\label{sec:induced-grav}

We have shown so far that the $\msu(n)$ gauge fields as well as 
scalar fields couple (almost-) covariantly to 
the effective metric $G^{ab}$.
However, we did not yet explain how the Einstein-Hilbert action
or some variation thereof should arise.
This appears to be difficult to achieve in the matrix-model framework,
where we can write down only traces of polynomials of the covariant
coordinates $X^a$. Moreover, adding any gauge-invariant term in the action
action would also affect the $\msu(n)$ sector which should describe
the Yang-Mills action.

We will argue that it is not necessary 
to add any further terms to the action,
rather the gravitational action 
arises automatically upon quantization.
The idea of induced gravity due to 
Sakharov \cite{Sakharov:1967pk} is crucial here; see e.g. 
\cite{Visser:2002ew} for a more recent discussion.
However, the term $\int \rho(y) \eta(y)$ in \eq{S-YM-effective} 
also plays an unexpected role.

Consider the quantization of the noncommutative gauge theory. 
The definition in terms of the 
matrix model actions \eq{YM-action-1} resp. \eq{scalar-action}
provides a clear quantization prescription 
via a (path) integral over the matrices $X^a$.
On the other hand, we can
use the description in terms of the classical
actions \eq{S-YM-effective} resp. \eq{S-scalar-eff}
at least for low energies, where the classical fields are coupled
covariantly to the effective metric $G^{ab}$. We can then use the 
well-known result that the one-loop effective action contains in particular
the Einstein-Hilbert action.

We briefly recall this general mechanism \cite{Visser:2002ew}: 
Consider e.g.
a scalar field with action 
$S[\Phi] = \int d^4 y\, \sqrt{\tilde g}\,\, 
\tilde g^{ab}\partial_a \Phi \partial_b\Phi$
coupled to some background metric $\tilde g$.
Upon quantization i.e. 
integration out $\phi$ up to a cutoff $\Lambda_{UV}$, the leading term
of the one-loop effective action
is essentially given by 
\be
S_{1-loop} \sim \int d^4 y \sqrt{\tilde g}\,
\(c_1 \Lambda_{UV}^4\, 
+ c_2 \Lambda_{UV}^2\, R[\tilde g]  + O(\log(\Lambda_{UV}))\)
\label{S-oneloop-seeley}
\ee
where $R[\tilde g]$ is the curvature scalar associated to $\tilde g$.
It involves the Seeley-de Witt coefficients  
determined by the kinetic terms
(see \cite{Gilkey:1995mj} \S 4.8). This is closely related to the 
spectral action principle \cite{Chamseddine:1996zu}, cf.  
\cite{Gayral:2004ww,Grosse:2007dm} for the Moyal-Weyl case. 

Our scalar action \eq{S-scalar-eff} differs from the generally covariant
form through a different power of $\det(\tilde g)$ in the measure
\eq{S-scalar-eff}.
This can be cast in the standard form by defining
\be
\tilde g_{ab} = e^{\sigma}\, G_{ab},
\qquad  e^{\sigma}\, = (\det G_{ab})^{-1/4}
\label{tilde-g}
\ee
with $\det \tilde g =1$, so that
\be
S[\Phi] =  c\int d^4 y\, (\det G_{ab})^{1/4}\, G^{ab}\partial_a \Phi \partial_b\Phi 
= c'  \int d^4 y\, \sqrt{\tilde g}\,\, \tilde g^{ab}\partial_a \Phi \partial_b\Phi .
\label{scalar-action-rescaled}
\ee
 This reflects the  invariance
of \eq{S-scalar-eff} under Weyl scaling. The curvature scalar of
$\tilde g_{ab}$ is related to the one for $G_{ab}$ by 
\be
R[\tilde g] 
= e^{-\sigma}\,\(R[G] - 3 \Delta_G \sigma - \frac 32\,
G^{ab} \partial_a\sigma \partial_b \sigma\)
%= e^{-\sigma}\,\(R[G] - \frac 32\,G^{ab} \partial_a\sigma \partial_b \sigma\)
\ee
where $e^{-\sigma} = \det(G)^{1/4}$ is somewhat reminiscent of a dilaton, and
\be
\Delta_G \sigma = \nabla^a_G \partial_a \sigma
= G^{ab} \partial_a \partial_b \sigma - \Gamma^c \partial_c \sigma \, .
\ee
Therefore \eq{scalar-action-rescaled} induces in particular the term
\be
S_{1-loop} \sim \int d^4 y \,
\det(G_{ab})^{1/4}\(R[G] -  3 \Delta_G \sigma-\frac 32\,G^{ab} \partial_a\sigma \partial_b \sigma\)\Lambda_{eff}^2
\label{S-oneloop-scalar}
\ee
at one-loop. 
This is just an indication of what should be expected from
a more detailed analysis. 
The $\msu(n)$ gauge fields will also induce at one loop 
terms similar to \eq{S-oneloop-scalar}.

\paragraph{UV/IR mixing and gravity.}

It is well-known that the quantization of noncommutative field theory
leads to the so-called UV/IR mixing 
\cite{Minwalla:1999px,Hayakawa:1999zf,Matusis:2000jf}. 
This means in particular that the effective action contains 
new divergent 
terms with momentum dependence $\sim\frac 1{(\theta p)^2}$, 
which are singular in the infrared and  not contained
in the bare action. This holds both for gauge fields and matter fields.
Remarkably, the UV/IR mixing for gauge fields is restricted to the 
trace-$\mmu(1)$ sector, at least for one loop.

Our result sheds new light on this phenomenon. We have argued 
using the semiclassical description that  NC gauge theory
induces upon quantization the Einstein-Hilbert 
action \eq{S-oneloop-seeley}
for the effective metric $G_{ab}$, which is a function 
of the $\mmu(1)$ gauge fields only, with divergent coefficients. 
Since these terms are not contained in the bare action,
the model  should {\em not}
be naively renormalizable as a pure Yang-Mills gauge theory,
and should have {\em new} divergences 
in the trace-$\mmu(1)$ sector (and only there) at one loop.
The momentum dependence of the scalar curvature $R$ \eq{R-wave},
valid for $k \ll \Lambda_{NC}$,
may well be responsible for the 
observed IR singularities in the naive $\mmu(1)$ point of view.
This shows that the essential features of the UV/IR mixing 
fit perfectly in our scenario and are 
in fact very welcome here. 

It remains to be seen 
how much this rough picture can be substantiated.
All of this underscores the importance of finite 
versions of NC gauge theory such as \cite{Behr:2005wp}
which are now understood as models of Euclidean quantum gravity, and of 
IR-modified versions such as \cite{Blaschke:2007vc} which might suppress
the gravitational sector.

Furthermore, recall that 
in the conventional framework, a major problem of 
induced gravity is that it induces huge 
cosmological constants. This problem is not expected to arise here, 
because the class of available metrics is restricted; 
note that the term $\int d^4 y \sqrt{\tilde g}\, \Lambda_{UV}^4$ 
in \eq{S-oneloop-seeley} is trivial here because $\sqrt{\tilde g}\equiv 1$.
Also, the term $\int d^4 y\, \rho(y) \eta(y)$ 
in \eq{S-YM-effective} does {\em not} play the role of the
cosmological ``constant'', rather it leads to the vacuum equations
of motion of gravity. These are the equations of motion for the 
$\mmu(1)$ degrees of freedom $Y^a$ for $\cF^{ab}=0=\Phi$, which 
are obtained  easily from \eq{eom-0}
\be
G^{ac} \partial_c \, \theta^{-1}_{ab}(y) =0 \,.
\label{eom-tree}
\ee
This will imply $R_{ab} \sim 0$ in the linearized case \eq{ricci-flat}. 
Furthermore, stability of Euclidean NC spaces with similar actions as 
the ones considered here 
is rather obvious by construction \cite{Behr:2005wp,Steinacker:2003sd} 
and has been verified 
numerically in \cite{Dou:2007in,Azuma:2004qe}, 
while geometrical phase transitions do  occur. 
Moreover, flat space \eq{moyal-weyl} 
remains to be a solutions even at one loop.
It therefore seems quite plausible that the picture of gravity emerging
from NC gauge theory may shed new light on  
the cosmological constant problem.

It remains to clarify the physical meaning of the 
metric $G_{ab}$ and possible rescaling with $e^{\sigma}$,
which is related to $\Lambda_{NC}$ via \eq{rho}.
Furthermore, the precise form of the gravitational 
equations of motion should be determined. 
We will show that at least for small fluctuations of flat
Minkowski space, the resulting gravity theory appears to be 
a physically acceptable modification
of Einstein gravity.

\paragraph{Relation to previous work on Matrix models and  M(atrix) theory.}

There is a large body of literature on 
Matrix-model formulations of string resp. M(atrix) theory.
In particular, the IKKT for IIB string theory
\cite{Ishibashi:1996xs} is essentially a 10-dimensional supersymmetric
version of the 4-dimensional model under consideration here,
while the BFSS model \cite{Banks:1996vh} for M-theory 
includes an extra ``time'' dependence. 
The identification of gravity in these matrix models is of particular
interest, and has been studied in 
a number of papers including \cite{Banks:1996vh,Kabat:1997sa,
Ishibashi:1996xs,Bigatti:1997jy,Aoki:1999vr,Ishibashi:2000hh,
Kimura:2000ur,Kitazawa:2006pj}. 
What is typically considered are 
interactions of separated ``D-objects'', represented 
by block-matrices. A gravitational interaction is then generated 
at one loop, i.e. by integrating out off-diagonal blocks,
reproducing leading effects of D=10 (super)gravity. 
However, there is also strong evidence for 
D=4 graviton propagators for D=4  D-brane solutions
\cite{Ishibashi:2000hh,Kitazawa:2006pj} of this matrix model,
which is quite directly related to the present context.
For other aspects see also \cite{Okawa:2000sh,Dhar:2001ba,Liu:2001ps}.
Nevertheless, an explicit identification of the associated geometries 
within such matrix models and 
its relation with gravity has not been obtained in the literature.

The relation with our approach is as follows. In stringy 
language, we consider a 
single given NC background (a 4-brane, say), and obtain an 
explicit metric and effective field theory. While 
these brane-solutions to the matrix models are typically
considered as flat (or highly symmetric), we point
out that they do contain nontrivial metrics and geometry through
their $U(1)$ sector.
In a higher-dimensional version, this should  
also shed new light on gravity in M(atrix) theory.
In agreement with previous work, 
one-loop effects are found to be 
crucial to obtain the gravitational action.

\subsection{Geometry, gravitational waves and $\mmu(1)$ gauge fields}
\label{sec:flat-expansion}

In this section we study in more detail the class of geometries 
available from \eq{effective-metric}. In particular, we
consider the case of small fluctuations around a
flat background $\R^4_\theta$ with generators
$\obar Y^a$. This will also clarify the 
relation with the conventional 
interpretation in terms of $\mmu(1)$ gauge fields on 
the canonical quantum plane $\R^4_\theta$.

An arbitrary $\mmu(1)$ component of $X^a$ in 
\eq{gaugefield-splitting} can be written as 
\be
Y^a = \obar Y^a + \obar \theta^{ab}\, A_b^0
\ee
so that
\be
\theta^{ab}(Y) = [Y^a,Y^b] =  \obar \theta^{ab} 
+  \obar \theta^{ac}\obar \theta^{bd}\, F^0_{cd}
\ee
where
$F^0_{cd} = \partial_c A_d^0 - \partial_d A^0_c + [A^0_c,A^0_d] $
is the abelian field strength on $\R^4_\theta$.
Therefore the induced metric can be written in terms of the $\mmu(1)$ 
gauge fields as 
\be
G^{ab} = -\theta^{ac} g_{cd} \theta^{b d} 
 = -(\obar \theta^{ac} + \obar \theta^{ae}\obar \theta^{ch}\, F^0_{eh}) 
(\obar \theta^{bd} + \obar \theta^{bf}\obar \theta^{dg}\,
F^0_{fg})g_{cd} \, .
\label{G-U1}
\ee
Consider first the case of 
2 dimensions. Then $\obar\theta^{ab} = \varepsilon_{ab} \obar\theta$ and
$F^0_{ab}(y) = \varepsilon_{ab} f(y)$, therefore
\be
G^{ab}_{(2D)}(y) = - g_{ab}\, \obar \theta^2 (1-  \obar\theta f(y))^2 .
\label{2Dmetric}
\ee
Since $g_{ab}$ is a constant diagonal matrix, 
the metric is obtained automatically in isothermal coordinates, and 
the $y$-dependence of the metric is given by the $y$-dependence of the
$\mmu(1)$ scalar field strength. 
The latter is an arbitrary function off-shell. Therefore 
the metric $G^{ab}_{(2D)}$ describes indeed the most general
metric in 2 dimensions with non-vanishing curvature, 
in isothermal ``gauge-fixing''.

In 4 dimensions, we certainly cannot obtain the most general 
geometry from the degrees of freedom of a $\mmu(1)$ gauge
field. However, we will show that one does obtain a class of metrics
which is sufficient to describe the 
physical (``on-shell'') degrees of freedom of gravity,
more precisely gravitational waves and the Newtonian limit for an 
arbitrary mass distribution.  

As a first check, note that gravitational waves have 2 physical 
degrees of freedom (helicities), as much as $\mmu(1)$ gauge fields. 
We should therefore verify whether \eq{G-U1} contains indeed
the 2 physical on-shell degrees of freedom of gravitational
waves on Minkowski space.
This was answered positively already in \cite{Rivelles:2002ez}
to leading order in $\obar\theta^{ab}$, and is reviewed below for
convenience. 
It strongly supports the physical viability of realizing gravity in this manner.

\paragraph{Gravitational waves on a flat background.}

Consider small fluctuations of the metric \eq{G-U1} 
around the metric for $\R^4_{\obar\theta}$
\be
\obar g^{ab} := -\obar \theta^{ac}\,\obar \theta^{bd}g_{cd} \, ,
\ee
which is indeed flat.
Keeping only the leading terms, \eq{G-U1} simplifies as
\bea
G^{ab} &=& \(\obar g^{ab} 
+ \obar g^{ad}\obar \theta^{bf}\, F^0_{df}
+ \obar g^{bd}\obar \theta^{af}\, F^0_{df} \) \,\, 
+ O(\obar g \obar \theta^2) \,.
\label{G-gauge-relation}
\eea
This can be considered as metric
fluctuations $G^{ab} = \obar g^{ab} - h^{ab}$ 
on flat Minkowski (or Euclidean) space, leading to
gravitational waves determined by 
\be
h^{ab} = -\obar g^{ad}\obar \theta^{bf}\, F^0_{df}
- \obar g^{bd}\obar \theta^{af}\, F^0_{df} \, .
\label{flat-grav-wave}
\ee
For the inverse metric $G_{ab} = \obar g_{ab} + h_{ab}$ this implies
\be
h_{ab} =  \obar g_{bb'}\obar \theta^{b'f}\, F^0_{fa}
+ \obar g_{aa'}\obar \theta^{a'f}\, F^0_{fb} 
\label{flat-grav-wave-2}
\ee
to leading order. This is essentially 
the metric obtained by Rivelles \cite{Rivelles:2002ez},
up to a trace contribution which arises here 
from the density $\rho(y)$ \eq{rho}.
Therefore the linearized picture in \cite{Rivelles:2002ez} 
is recovered here in a complete framework with nontrivial geometry. 
The linearized Ricci tensor is found to be
\bea
R_{ab} &=& \partial^c \partial_{(b} h_{a)c}
   - \frac 12 \partial^c \partial_c h_{ab} 
   - \frac 12 \partial_a \partial_b h \nn\\
&=& \frac 12\( \obar \theta^{cf}\, \partial_a\partial_b  F^0_{cf}
- {\obar \theta_a}^{f}\,\partial^c \partial_{f} F^0_{cb}
-  {\obar\theta_b}^{f}\, \partial^c\partial_{f} F^0_{ca} \) 
\eea
where indices are raised and lowered with $\obar g$, 
\be
h = h_{ab} \obar g^{ab} = 2 \obar \theta^{af}\, F^0_{fa}\, ,
\label{h-wave}
\ee
and
\be
R =  \obar\theta^{af}\, \partial^c \partial_c F^0_{af} \, .
\label{R-wave}
\ee
On the other hand, the linearized Ricci tensor for the unimodular
metric $\tilde g_{ab}$ \eq{tilde-g} resp. the traceless graviton
$\tilde h_{ab} = h_{ab} -\frac 14 \obar g_{ab} h$ is given by
\be
R_{ab}[\tilde g] =
  -\frac 12\({\obar \theta_a}^{f}\,\partial^c \partial_{f} F^0_{cb}
+  {\obar\theta_b}^{f}\, \partial^c\partial_{f}  F^0_{ca}
+\frac 12 \obar g_{ab}\partial^c \partial_c F_{de}\theta^{de}\)
\label{ricci-unimodular}
\ee
This agrees with the results of \cite{Rivelles:2002ez}.
Now consider the tree-level vacuum equations of motion \eq{eom-tree},
which in the present context amount to 
$\partial^a F_{ab}^0=0 = \partial^c\partial_{c}  F^0_{ab}$
up to possibly corrections of order $\theta$,
i.e. the vacuum Maxwell equations for the flat metric $\obar g_{ab}$.
As pointed out in \cite{Rivelles:2002ez}, this implies that
the vacuum geometries are Ricci-flat,
\be
R_{ab}[\tilde g] = 0 + O(\theta^2), %\qquad R = 0 + O(\theta^2)
\label{ricci-flat}
\ee
while the general curvature tensor $R_{abcd}$ is first order in
$\theta$ and does not vanish. This
shows that the effective metric does contain the 2 physical 
degrees of freedom (helicities) of
gravitational waves. It is quite remarkable that this is obtained 
at the tree level,
without invoking the mechanism of induced gravity 
in section \ref{sec:induced-grav}.

For completeness, we check that the Riemann tensor 
for  plane waves is non-zero. To do this the following 
form of the metric fluctuations \eq{flat-grav-wave-2} is more convenient
\bea
h_{ab} &=&  {\obar \theta_b}^{f}\, \partial_f A_a^0 
+ {\obar\theta_a}^{f}\,  \partial_f A_b^0
 - (\partial_a A_f^0 \,{\obar \theta_b}^{f}\,
 + \partial_b A_f^0\, {\obar \theta_a}^{f}) \, \nn\\
 &\cong& {\obar \theta_b}^{f}\, \partial_f A_a^0 
+ {\obar \theta_a}^{f}\,  \partial_f A_b^0
\label{flat-grav-wave-3}
\eea
since the term in brackets 
has the form $\partial_a \xi_b + \partial_b \xi_a$
of an infinitesimal diffeomorphism and therefore can be dropped.
Incidentally, observe that the $\mmu(1)$ 
gauge transformations act as 
$A_a^0 \to A_a^0 + \partial_a \lambda(x)$ in the commutative limit, which leaves
 $h_{ab}$ invariant; 
%% new in V2
however,
they do act as symplectomorphism to order $\theta$, as discussed in 
section \ref{sec:symplectomorphisms}.
Now consider plane-wave configurations  
\be
A^0_a = E_a\, e^{i k x}
\ee
with
\be
h_{ab} =  i ({\obar \theta_b}^{f} k_f E_a + {\obar \theta_a}^{f} k_f E_b)\, .
\ee
Using
\be
\Gamma^c_{ab} = \frac 12 \obar g^{cd} \(\partial_a h_{bd} + \partial_b h_{ad}
 - \partial_d h_{ab}\) ,
\ee
the linearized curvature tensor is 
\be
{R_{abc}}^d 
= -i\frac 12 \Big(
  (k_c  \obar \theta^{df} - k^d  {\obar \theta_c}^{f}) k_f (k_b E_a - k_a E_b) 
+ (k_b {\obar\theta_a}^{f} - k_a {\obar\theta_b}^{f}) k_f (k_c E^d -  k^d E_c)\Big) \nn\\
\ee
which is $O(\theta)$ and does not vanish even on-shell.

This analysis suggests in particular that
gravitons should be interpreted as NC 
Goldstone bosons for the spontaneously broken translational
invariance of $X^a\to X^a + c^a$, 
and gauge bosons as their nonabelian cousins.

\subsection{Connection and curvature, examples}
\label{sec:examples}

The Christoffel symbols obtained from the metric $G^{ab}$ 
for general $\theta^{ab}(y)$are
\be
\Gamma^c_{ab} = \frac 12 G^{cd} \(\partial_a G_{bd} + \partial_b G_{ad}
 - \partial_d G_{ab}\) 
\ee
which using the Jacobi identity for $\theta^{-1}_{ab}$ can be written as 
\be
\Gamma^c_{ab}
=\frac 12 \(\theta^{cf}\partial_a \theta^{-1}_{bf} 
+\theta^{cf}\partial_b \theta^{-1}_{af}
 + G^{cd} (\theta^{-1}_{bf} g^{ff'} \partial_{f'} \theta^{-1}_{ad}
 +\theta^{-1}_{af}g^{ff'}\partial_{f'} \theta^{-1}_{bd})\) \,.
\label{christoffel-explicit}
\ee
The curvature is given as usual by
\be
{R_{abc}}^d = \partial_b \Gamma^{d}_{ac} - \partial_a\Gamma^{d}_{bc}
 + \Gamma^{e}_{ac}\Gamma^{d}_{eb} - \Gamma^{e}_{bc}\Gamma^{d}_{ea}\, .
\ee
Inserting \eq{christoffel-explicit} does not provide very illuminating
expressions.
Note that $\theta^{ab}(y)$ is in general not covariantly constant, 
even though $G^{ab}$ is.

We illustrate the nontrivial 
geometries emerging from NC spaces with a few  examples.

\paragraph{Manin plane.}

Consider the Manin plane 
\be
x y = q y x
\ee
with $|q|=1$ and hermitian generators $x,y$. The underlying Poisson
structure is 
\be
\{x,y\} = -i(q-q^{-1})\, xy =: -i\theta(x,y)
\ee
so that the effective metric induced by the matrix model with
background metric $g_{ab} = \delta_{ab}$ resp. $g_{ab} = \eta_{ab}$
would be
\be
ds^2 = -(q-q^{-1})^2\, x^2 y^2 (dx^2 \pm dy^2) \,.
\ee
However, keep in mind that the Manin plane might be obtained more
naturally from
a different matrix model with different background metric $g_{ab}$,
with different $G_{ab}$.

\paragraph{Newtonian limit.}

The Newtonian limit of general relativity corresponds to static metric
perturbations of the form 
\be
ds^2 = -c^2 dt^2\Big(1+\frac {2U}{c^2}\Big) 
+  d \vec x^2 \Big(1+O(\frac 1{c^2})\Big)
\label{newton-metric}
\ee
where $\Delta_{(3)} U = 4\pi G\rho$ and $\rho$ is the mass density.
We can indeed obtain such metrics for arbitrary static $\rho$, 
as shown in Appendix B \eq{h-newton}.
Therefore the class of metrics $G_{ab}$ \eq{effective-metric}
does contain the required degrees of freedom to describe a physically
reasonable gravity theory. In fact, the degrees of freedom 
for $G_{ab}$ are {\em precisely} those required to describe an arbitrary
mass distribution. This gravity theory is therefore very economical. 
The Planck length is identified with $\L_{NC}^{-1}$ on dimensional
grounds, or via \eq{U-def} which gives $G \sim \theta$ in appropriate units.

If we us the vacuum equations of motion \eq{eom-tree}
which amounts to $\partial^c F_{cb} =0$ resp. $R_{ab}=0$
as discussed above,
then \eq{h-newton} leads to
\be
ds^2 = -c^2 dt^2\Big(1+\frac {2U}{c^2}\Big) 
+  d \vec x^2 \Big(1-\frac {2U}{c^2}\Big)
\ee
to leading order, as in general relativity. 
Therefore the leading
corrections of general relativity over 
Newtonian gravity should be reproduced here.

\paragraph{Schwarzschild metric, rescaled.}

The Schwarzschild metric  can be written in Kruskal coordinates as
\be
ds^2 = r^2 (d \vartheta^2 + \sin^2(\vartheta) d\varphi^2)
 + \frac 4{r}\, e^{-r} (du^2 - dv^2) 
\label{kruskal}
\ee
where $u+v = \sqrt{r-1}\, e^{(r+t)/2}, \,\,  u-v = \sqrt{r-1}\,
e^{(r-t)/2}$ and thus $u^2-v^2 = (r-1) e^r$.
This can be written as
\be
G_{ab} = r^2 \tilde G_{ab}
= r^2 \theta^{-1}_{a a'} \theta^{-1}_{b b'}\, \eta^{a'  b'} 
\ee
which  almost the desired form (except for the overall scaling 
factor $r^2$) for the symplectic form 
\be
\theta^{-1}_{ab} dx^a \wedge dx^b =  \sin(\vartheta) d\vartheta d\varphi 
+ \frac 2{r^{3/2}}\, e^{-r/2} du \wedge dv \,.
\ee
 Note that the density factor 
$e^{\sigma} = (\det \tilde G)^{1/4} = \frac{2}{r^{3/2}}\, e^{-r/2}$
 is a function of $r$ only, so that the 
\eq{kruskal} is indeed obtained by rescaling with a function of
$\sigma$ only. 
%We will not discuss this any further here since 
%the role of $e^{\sigma}$ and the
%gravitational action is not clear at this point. 
% new V2
The $(r,t)$ - part of the metric can easily be generalized
as in \eq{2Dmetric}. While this illustrates the nontrivial nature of
metrics of the form \eq{effective-metric},
it turns out that this ansatz does not lead to the desired
Schwarzschild-like solution, rather a different ansatz must be used;
this will be described elsewhere.

\subsection{Coordinates, gauge invariance and symplectomorphisms}
\label{sec:symplectomorphisms}

From a semiclassical point of view,
NC gauge theory provides 2 geometrical structures: 
1) a Poisson structure $\theta^{ab}(x)$
and 2) a ``background'' (closed string) metric $g_{ab}$, which is used
to contract the indices of the covariant coordinates.
We assume here that $g_{ab}$ is flat. 
There are accordingly 2 special coordinate systems:
\begin{enumerate}
\item
Darboux coordinates where $\theta^{ab}$ is constant. Then of course
the background metric $g_{ab}(x)$ is {\em not} given by 
$\delta_{ab}$ or $\eta_{ab}$, but it is still flat. 
\item Cartesian coordinates w.r.t. the background metric  $g_{ab}$.
Then $\theta^{ab}(y)$ is not constant. 
These are the $y^a$ coordinates used in the present paper.
\end{enumerate}
Observe that $G_{ab}$ is flat if the two coincide, thus NC gravity
results in some sense from a ``strain'' 
between Darboux- and $g$-flat coordinates.

Now consider the gauge symmetries.
The  matrix-model action \eq{YM-action-1} is invariant under
the NC gauge transformations \eq{NCgaugetrafo}. While their 
$\msu(n)$ components are clearly the $\msu(n)$ gauge
transformations of the effective action \eq{S-YM-effective}, the
role of the local $\mmu(1)$ transformations is less obvious.
It is well-known (see e.g. \cite{Lizzi:2001nd})
that $\mmu(1)$ gauge transformations in the NC
case act naturally as symplectomorphisms on the 
Poisson manifold  $\cM$, leaving $\theta^{ab}(y)$ invariant. 
To see this, consider the gauge transformation of
a scalar function $\phi(y) \in \cA$:
\be
\phi \to \phi' = U \phi U^{-1} 
\ee
or infinitesimally
\be
\phi \to \phi' = \phi + i[\Lambda, \phi] 
\label{infintes-sympl} 
\ee 
for $U = e^{i \varepsilon\Lambda}$. 
The semi-classical version of this action
is $\phi(y) \to \phi'(y) = \phi(y) + \{\Lambda(y), \phi(y)\}$, 
which generates the Hamiltonian 
flow with generator $\Lambda(y)$ w.r.t. the 
Poisson structure $\theta^{ab}(y)$. 
Therefore $\mmu(1)$ gauge transformations are naturally interpreted as
quantization of the 
action of the group $\rm{Symp}(\cM)$ of symplectomorphisms on
$\cM$. Due to Liouvilles theorem, $\rm{Symp}(\cM)$ is a (proper) 
subgroup of the group of
volume-preserving diffeomorphism. 

Now consider the covariant coordinates $X^\a$, which transform as 
\be
X^a \to {X^a}' = U^{-1} X^a U .
\label{X-symplecto}
\ee
According to the above discussion, this can be interpreted 
for the $\mmu(1)$ sector as
transformation of the embedding function
$X^a: \cM \hookrightarrow \R^4$ under (quantized)
$\rm{Symp}(\cM)$. However here $\rm{Symp}(\cM)$
does {\em not} act on any indices of e.g. nonabelian gauge fields, 
unlike the standard action of 
diffeomorphisms. 
Nevertheless, since the action is written in
terms of classical field strength
tensors with all indices properly contracted, the 
classical action appears to be general covariant.
This is only apparent, however, since   
$g_{ab}$ is a fixed  background metric: The exact
invariance group must preserve $\rho$ and $\eta(y)$, which probably 
reduces it to $\rm{Symp}(\cM)$.

The role of NC gauge transformations and diffeomorphisms certainly 
deserves further investigations, see also 
\cite{Yang:2004vd,Azuma:2001re} for related discussion. It remains to be
seen whether the generalized notions of symmetry 
developed in \cite{Aschieri:2005zs} 
are applicable in the context of matrix models.

\section{Remarks on the quantization}

The great virtue of matrix models such as \eq{YM-action-1}
is that there is a clear concept of quantization, defined by  
integrating over the space of matrices. This has been
extremely successful for single-matrix models, and was elaborated
in the context of NC gauge theory 
to some extent \cite{Steinacker:2003sd}. 
Combined with the results of the present paper,  
this leads to the hope that \eq{YM-action-1} may provide a good
definition of quantum gravity. 
The limit $N \to \infty$ of course remains to be 
a highly nontrivial issue related to renormalizability. 
On the other hand, the finite-dimensional 
matrix-models for compact ``fuzzy'' quantum spaces 
such as \cite{Behr:2005wp} are thus
candidates for a regularized (Euclidean) gravity theory.

Furthermore, recall from section 
\ref{sec:examples} that our model of 
NC gravity contains only the minimal
degrees of freedom required to accomodate on-shell 
gravitational waves plus a mass distribution. 
In contrast, general relativity contains many additional
off-shell and gauge degrees of freedom, leading 
in particular to nontrivial gauge fixing issues upon quantization. 
Therefore the gravity theory obtained here should be 
better suited for quantization.

We support this conjecture with some observations. 
Due to gauge invariance \eq{NCgaugetrafo}, 
the effective action after quantization should
be given by similar types 
of matrix models, involving more complicated expressions of traces of
polynomials of the $X^a$. Due to translational invariance, they should
be expressible in terms of commutators, and therefore - in some given
vacuum - the same analysis as here should establish that they can be
interpreted as $\msu(n)$ gauge theory coupled to an effective
$G^{ab}$, to leading order. 
This suggests that there should be no 
disastrous UV/IR mixing effect, which has been absorbed by the choice
of geometric vacuum.

\section{Discussion}

The basic message of this paper is that 
gravity is an intrinsic part 
of the matrix-model formulation of NC gauge theory. 
These models describe a dynamical noncommutative
space, with metric determined by the general Poisson structure.
This leads to a separation of the 
gravity and gauge theory degrees of freedom.
Quantum spaces and gravity are seen as two aspects of the same thing.
Matrix models such as \eq{YM-action-1} thus
provide a simple class of models which should be suitable 
for quantizing gravity along with the other fields.
This clarifies 
the presence of gravity in string-theoretical matrix models
\cite{Banks:1996vh,Ishibashi:1996xs}, 
however the mechanism is more general and 
applies in particular to 4 dimensions, as elaborated here.
Also, the mechanism of spontaneous
generation of fuzzy extra dimensions \cite{Aschieri:2006uw}
can now be seen from the point of view of gravity.
We also point out that 
the gravitational action will be induced upon
quantization, which should explain and hopefully 
resolve the UV/IR mixing in NC gauge theory.

While the physical properties of the emerging gravity theory
are not yet worked out, 
the simplicity of the mechanism is certainly striking. 
There remains some freedom for modification of the
action, in particular via extra dimensions, 
but the mechanism seems to be quite rigid. In particular,
the restricted class of geometries strongly suggests that
the resulting gravity theory is different from
general relativity, but consistent with its low-energy limit. 
This realizes some of the ideas in
\cite{Rivelles:2002ez,Yang:2004vd,Banerjee:2004rs}, 
with the aim to understand gravity as an emergent phenomenon
of NC gauge theory in the commutative limit. 
It is also reminiscent to ideas in \cite{Madore:2000aq}, in the sense
that gravity is determined by noncommutativity 
i.e. the Poisson structure. On the other hand, 
this is different from other proposals \cite{Aschieri:2005zs}
which aim to define a deformed (noncommutative) 
version of general relativity. 

One may wonder how such a different interpretation of 
NC gauge theory is possible; after all, 
there seems to be nothing wrong with 
the ``old'' gauge theory point of view. 
From that perspective, what we have done is 
to perform a Seiberg-Witten map from constant $\obar \theta^{ab}$ to a
general $\theta^{ab}(y)$, to leading order in $\theta^{ab}(y)$ but
exact in $\delta \theta^{ab} = \theta^{ab}(y) - \obar\theta^{ab}(y)$. 
This ``eats up'' the $\mmu(1)$ gauge fields and moves them into the
metric $G^{ab}(y)$. 
In the conventional gauge theory point of view, $\delta \theta^{ab}$
is the $\mmu(1)$ field strength, which decouples from the $\msu(n)$
gauge degrees of freedom to leading order but cannot be disentangled
exactly. We determined the precise coupling between these $\mmu(1)$
and  $\msu(n)$ degrees of freedom, and showed that it should be
interpreted as gravitational coupling. This casts the 
basic observations of \cite{Rivelles:2002ez} in a complete
framework, generalized to notrivial geometries 
and nonabelian gauge fields. The basic idea of 
gravity emerging form NC gauge theory was also put forward in 
\cite{Yang:2004vd,Banerjee:2004rs}, in a somewhat different
approach without identifying the metric \eq{effective-metric}.

There are many further directions to explore. First, 
the main results of this paper also apply to 
dimension different from 4, 
and should generalize in particular to the case of 
NC ``submanifolds'' embedded in higher dimensions. 
Then the closed string metric 
$g_{ab}$ is the induced metric on the submanifold, and no longer
flat in general. Therefore the class of effective metrics
obtained in this case may be larger. 
Notice also that extra dimensions can be viewed as additional
(possibly interacting) scalars as in \eq{scalar-action-0}; a 
particularly interesting example would be 
the matrix model for $N=4$ NCSYM considered e.g. in 
\cite{Ishibashi:2000hh,Kitazawa:2006pj}.
Other types of matrix model actions should also be explored, 
such as DBI-like actions. 
Fermions should of course be included in these models, which
will be studied elsewhere. This will also
allow to study the relation with 
the framework of the spectral action \cite{Chamseddine:1996zu}.
The quantization and loop effects should be worked out.
Finally, it is of course essential to explore the physical viability
of this NC gravity.

\subsection*{Acknowledgments}

This paper is dedicated to the memory of Julius Wess, who took a
leading role in the development of this circle of ideas in physics,
and provided essential support and an open spirit.

I would like to thank in particular H. Grosse for 
discussions on various aspects of this work, as well as C-S. Chu,
J. Wess, and H.-S. Yang for
stimulating exchange on various aspects and related ideas. 
I am also grateful to M. Wohlgenannt 
for pointing out \cite{Rivelles:2002ez} to me.
This work was supported by the FWF Project P18657.

\section{Appendix A: Derivation of the effective action to 
leading order}

To shorten the notation we only consider the Euclidean case
$g_{ab} = \delta_{ab}$ here, and adopt a notation where repeated
indices are summed irrespective of their position; for example,
$\theta^{ab}\theta^{ab} \equiv \sum_{a,b}\theta^{ab}\theta^{ab}$.
The Minkowski case is obtained by obvious replacements.

Furthermore, we adopt the convention in this appendix 
to rise and lower indices with
$\theta^{ab}$
resp. $\theta^{-1}_{ab}$ rather than the metric, 
e.g. $A^a = \theta^{ab}\, A_b$.

\subsection*{Useful identities}

The ``commutative'' field strength is defined by
\bea
F^{ab} &=& \theta^{ac} \theta^{bd} F_{cd} 
=  \theta^{ac} \theta^{bd} (\partial_c A_d - \partial_d A_c) 
+ \theta^{ac} \theta^{bd}[A_c,A_d] \nn\\
&=& \theta^{bd}[Y^a, A_d] -\theta^{ac} [Y^b, A_c]  + \theta^{ac} \theta^{bd}[A_c,A_d] 
\label{F-explicit-2}
\eea
while we define the ``noncommutative'' field strength as
\bea
\cF^{ab} &=& [X^a,X^b] - \theta^{ab}
= [Y^a,\cA^b] - [Y^b,\cA^a] + [\cA^a,\cA^b] \nn\\
&=& [X^a,\cA^b] - [X^b,\cA^a] - [\cA^a,\cA^b] \, .
\eea
The leading terms are
\bea
\cF^{ab} &=& [Y^a,A_d \theta^{bd}] - [Y^b,A_d \theta^{ad}] 
  + [A^a,A^b] \nn\\
&=& F^{ab} + ([Y^a,\theta^{bd}] - [Y^b,\theta^{ad}])A_d 
 + [A_d \theta^{ad},A_e \theta^{be}] - \theta^{ad}\theta^{be}[A_d ,A_e]\nn\\
&=& F^{ab} - A_d[Y^d,\theta^{ab}] 
 + [A^a,A^b] - \theta^{aa'}\theta^{be'}[A_{a'} ,A_{e'}]
\label{F-identity}
\eea
up to corrections of order $O(\theta^3)$, hence
omitting $\cF^{ab}_{SW,2}$ here.

A useful identity is 
\be
2\theta^{ab}[Y^a,[Y^b,X]] 
= \theta^{ab}([Y^a,[Y^b,X]] - [Y^b,[Y^a,X]])
= \theta^{ab}[\theta^{ab},X]\, .
\label{theta-a-b-id}
\ee
A similar identity is the following:
\bea
 \theta^{ab} [Y^a,\theta^{cb}]
&=&  - \theta^{ab} [Y^c,\theta^{ba}]
- \theta^{ab} [Y^b,\theta^{ac}]  \nn\\
&=&  - \theta^{ab} [Y^c,\theta^{ba}]
+ \theta^{ab} [Y^a,\theta^{bc}]  \nn
\eea
therefore
\be
  \theta^{ab} [Y^a,\theta^{cb}] 
= \frac 12  \theta^{ab} [Y^c,\theta^{ab}] \, .
\label{yy-id}
\ee
In particular, 
\be
  \theta^{ab} A_c A_d [Y^d,[Y^a,\theta^{cb}]] 
= \frac 12  \theta^{ab} A_c A_d  [Y^d,[Y^c,\theta^{ab}]] \, .
\label{aayy-id}
\ee

\paragraph{Bianci identity and applications}

The noncommutative Bianci identity for $\cF$ 
is obtained from
\bea
[X^a,\cF^{bc}] + [X^b,\cF^{ca}] + [X^c,\cF^{ab}]
 &=& - [X^a,\theta^{bc}] - [X^b,\theta^{ca}] - [X^c,\theta^{ab}]\nn\\
&=& - [\cA^a,\theta^{bc}] - [\cA^b,\theta^{ca}] - [\cA^c,\theta^{ab}] \, .
\eea
Together with the antisymmetry of $\theta^{ab}$, it follows that
\bea
\theta^{ab} [X^a,\cF^{cb}] 
&=&  \theta^{ab} \(-[X^c,\cF^{ba}] - [X^b,\cF^{ac}] 
  - [\cA^a,\theta^{cb}] - [\cA^b,\theta^{ac}] - [\cA^c,\theta^{ba}] \)\nn\\
%&=& \theta^{ab}\(-[X^c,\cF^{ba}] + [X^a,\cF^{bc}] 
%- [\cA^a,\theta^{cb}] + [\cA^a,\theta^{bc}] - [\cA^c,\theta^{ba}]\)\nn\\
&=&  \theta^{ab} \(-[X^c,\cF^{ba}] - [X^a,\cF^{cb}] 
- [\cA^a,\theta^{cb}] - [\cA^a,\theta^{cb}] + [\cA^c,\theta^{ab}]\)
\eea
which  implies
\be
\theta^{ab} [X^a,\cF^{cb}] 
= \frac 12 \theta^{ab} ([X^c,\cF^{ab}]+[A^c,\theta^{ab}]) 
-  \theta^{ab}[A^a,\theta^{cb}] \, .
\label{thetadF-id}
\ee
Using $[Y^c,A^b] + [A^c,A^b] = \cF^{cb} + [Y^b,A^c]$ this gives
\bea
\theta^{ab} [X^a,[Y^c,A^b] + [A^c,A^b]]]
&=& \theta^{ab} [X^a,\cF^{cb}] 
+ \theta^{ab} [X^a,[Y^b,A^c]]  \nn\\
&=& \frac 12\theta^{ab} ([X^c,\cF^{ab}]+[A^c,\theta^{ab}])
+ \theta^{ab} [Y^a,[Y^b,A^c]] \nn\\
&& + \theta^{ab} [A^a,[Y^b,A^c]] 
- \theta^{ab}[A^a,\theta^{cb}] \nn\\
&=& \frac 12\theta^{ab} ([X^c,\cF^{ab}]+[A^c,\theta^{ab}])
+ \frac 12 \theta^{ab} [\theta^{ab},A^c] \nn\\
&& + \theta^{ab} [A^a,[Y^b,A^c]] 
- \theta^{ab}[A^a,\theta^{cb}] \nn\\
% &=& \frac 12\theta^{ab} [X^c,\cF^{ab}]
% - \theta^{ab} [Y^b,[A^c,A^a]] 
% - \theta^{ab} [A^c,[A^a,Y^b]] 
% - \theta^{ab}[A^a,\theta^{cb}] \nn\\
&=&\theta^{ab}\Big(\frac 12 [X^c,\cF^{ab}]
+ [Y^a,[A^c,A^b]] -  [A^c,[A^a,Y^b]] - [A^a,\theta^{cb}] \Big)\nn
\eea
so that
\bea
\theta^{ab} [X^a,[Y^c,A^b]]
&=&\frac 12\theta^{ab} [X^c,\cF^{ab}] 
- \theta^{ab} [A^c,[A^a,Y^b]] 
- \theta^{ab} [A^a,[A^c,A^b]]
- \theta^{ab}[A^a,\theta^{cb}] \nn
\eea
which using
\bea
\theta^{ab} [A^a,[A^c,A^b]]
%&=& - \theta^{ab} [A^c,[A^b,A^a]]- \theta^{ab}[A^b,[A^a,A^c]] \nn\\
&=& - \theta^{ab} [A^c,[A^b,A^a]]+ \theta^{ab}[A^a,[A^b,A^c]] 
\eea
thus
\be
\theta^{ab} [A^a,[A^c,A^b]] = \frac 12 \theta^{ab} [A^c,[A^a,A^b]]
\ee
gives
\bea
\theta^{ab} [X^a,[Y^c,A^b]]
&=&\frac 12\theta^{ab} [X^c,\cF^{ab}] 
- \theta^{ab} [A^c,[A^a,Y^b]] 
- \frac 12 \theta^{ab} [A^c,[A^a,A^b]]
- \theta^{ab}[A^a,\theta^{cb}] \nn\\
&=&\frac 12\theta^{ab} [X^c,\cF^{ab}] 
- \frac 12\theta^{ab} [A^c,F^{ab}] 
- \theta^{ab}[A^a,\theta^{cb}] \, .
\label{acd-id-3}
\eea

\paragraph{Other useful relations}

Here we collect some identities which hold up to 
some required order or $\theta$.

Let us introduce the notation
\be
[Y_a,f] := \theta^{-1}_{ab}\, [Y^b,f] = \partial_a f \quad + O(\theta)
\ee
which allows to write
\be
 [Y^a,f] [Y_a,g] =  \theta^{ab} \partial_b f \partial_a g = - i\{f,g\}
\quad + O(\theta^2)
\ee
to leading order, which in the abelian case 
coincides with $-[f,g] + O(\theta^2)$. This gives
\bea
 \theta^{ab} [X^a,A_c] (F^{cb} + [Y^c, A^b])
&=& \theta^{ab}({F^a}_c + [Y_c,A_e]\theta^{ae}) (F^{cb} + [Y^c,A^b]) \nn\\
&=&  \theta^{ab} ({F^a}_c F^{cb} + [Y_c,A_e] \theta^{ae}[Y^c,A^b]
   + {F^a}_c [Y^c,A^b] + F^{cb} [Y_c,A_e] \theta^{ae})  \nn\\
&=&  \theta^{ab} ({F^a}_c F^{cb} + \theta^{ae} i\{A_e,A^b\}
   - F^{cb} A_e[Y_c,\theta^{ae}]) \nn\\
&=& \theta^{ab}F^{ad}\theta^{-1}_{dc} F^{bc} 
+  \theta^{ab} F^{ad} \theta^{-1}_{dc} A_e [Y^c,\theta^{eb}] 
 + \theta^{ab}\theta^{ae} i\{A_e, A^b\} 
\label{FF-id-2-1}
\eea
up to $O(\theta^4)$.
Similarly, one finds
\bea
\theta^{ab} (A_d [X^a,A_c] [Y^c,\theta^{db}] + F^{ad} \theta^{-1}_{dc}A_e [Y^c,\theta^{eb}])
&=&  \theta^{ab} A_d ([X^a,A_c] - {F^a}_c ) [Y^c,\theta^{db}]) \nn\\
&=&  \theta^{ab} A_d [Y_c,A_e]\theta^{ae}  [Y^c,\theta^{db}] \nn\\
&\sim&   \theta^{ab}  \theta^{ae} A_d [A_e,\theta^{db}] \, .
\label{FA-comb}
\eea
To evaluate the contributions cubic in $A$, we will need
\bea
Tr \theta^{ab} \theta^{ab} A_d [\cF^{cd},A_c] 
&=&- Tr \theta^{ab} \theta^{ab} [A_d,A_c] \cF^{cd} \nn\\
&=& - Tr \theta^{ab} \theta^{ab} [A_d,A_c]([Y^c,A^d] - [Y^d,A^c] + [A^c,A^d]) \nn\\
&=& - Tr \theta^{ab} \theta^{ab} [A_d,A_c](2[Y^c,A^d]+ [A^c,A^d]) \, .
\label{AFA}
\eea
The first term gives
\bea
Tr \theta^{ab} \theta^{ab} [A_d,A_c][Y^c,A^d] &=&
Tr -\theta^{ab} \theta^{ab} A_d [[Y^c,A^d],A_c] \nn\\
&=& Tr \theta^{ab} \theta^{ab} A_d ([[A^d,A_c],Y^c] + [[A_c,Y^c],A^d])\nn\\
&=& Tr \theta^{ab} \theta^{ab} (-[A_d,Y^c] [A^d,A_c] - [A_d,A^d] [A_c,Y^c])\, .
\label{AAYA}
\eea 
To proceed, consider
\bea
Tr \theta^{ab} \theta^{ab} [A^d,Y^c] [A_d,A_c] 
&=& Tr \theta^{ab} \theta^{ab} [\theta^{de} A_e,Y^c] [A_d,A_c] \nn\\
&=&  Tr \theta^{ab} \theta^{ab} ([\theta^{de},Y^c] A_e + [ A_e,Y^c]\theta^{de})[A_d,A_c]\nn\\
&=&  Tr \theta^{ab} \theta^{ab} ([\theta^{de},Y^c] A_e [A_d,A_c]- [A_e,Y^c][A^e,A_c])\nn\\
&=& - Tr \theta^{ab} \theta^{ab}[A_e,Y^c][A^e,A_c]
\eea
dropping terms of order $O(\theta^5)$ and
using \eq{AAAY-id} below, which can be obtained by
considering
\bea
Tr\theta^{ab}\theta^{ab} [A_c,A_d]  A_e [Y^e,\theta^{cd}]
&=& Tr\theta^{ab}\theta^{ab} [A_c, A_d]  A_e (-[Y^c,\theta^{de}]+[Y^d,\theta^{ce}]) \nn\\
&=& -2 Tr\theta^{ab}\theta^{ab} [A_c, A_d]  A_e [Y^c,\theta^{de}] \nn\\
&=& - 2 Tr\theta^{ab}\theta^{ab} [A_d, A_e] A_c  [Y^c,\theta^{de}] ,
\label{AAA-triple}
\eea
routinely dropping terms of the type $Tr \theta^4\, f(x) [A_a,A_b]_{n.a.}$
under the trace.
This implies that
\be
Tr\theta^{ab}\theta^{ab} [A_c,A_d]  A_e [Y^e,\theta^{cd}] =
Tr\theta^{ab}\theta^{ab} [A_c, A_d]  A_e [Y^c,\theta^{de}] = 0 \, .
\label{AAAY-id}
\ee
Therefore \eq{AAYA} gives
\be
Tr \theta^{ab} \theta^{ab} [A_d,A_c][Y^c,A^d] =
Tr \theta^{ab} \theta^{ab} (-[Y^c,A^d] [A_d,A_c] - [A_d,A^d][A_c,Y^c])
\ee
which implies
\be
Tr \theta^{ab} \theta^{ab} [A_d,A_c][Y^c,A^d] 
= - \frac 12 Tr \theta^{ab} \theta^{ab}[A^d,A_d][Y^c,A_c] \, .
\ee
Similarly,
\bea
- Tr \theta^{ab} \theta^{ab} [A_d,A_c][A^c,A^d]
&=& Tr \theta^{ab} \theta^{ab} A_c[A_d,[A^c,A^d]] \nn\\
&=&  Tr \theta^{ab} \theta^{ab} A_c (-[A^c,[A^d,A_d]] - [A^d,[A_d,A^c]] )\nn\\
&=&  Tr \theta^{ab} \theta^{ab} ([A^c,A_c] [A^d,A_d] + [A^d,A_c]  [A_d,A^c])\nn\\
&=&  Tr \theta^{ab} \theta^{ab} ([A^c,A_c] [A^d,A_d] - [A_d,A_c]  [A^d,A^c])
\eea
implies
\be
 Tr \theta^{ab} \theta^{ab} [A_d,A_c][A^c,A^d]
= -\frac 12 Tr \theta^{ab} \theta^{ab}[A^c,A_c] [A^d,A_d] \,.
\ee
Putting this together, \eq{AFA} can be written as
\be
Tr \theta^{ab} \theta^{ab} A_d [\cF^{cd},A_c] 
= \frac 12 Tr \theta^{ab} \theta^{ab}(2[A^d,A_d][Y^c,A_c] +
[A^c,A_c] [A^d,A_d])  \,.
\label{AFA-id-2}
\ee

\subsection*{Evaluation of the  contributions}

\subsubsection*{second-order Seiberg-Witten contribution}

Let us write the second-order 
Seiberg-Witten contributions \eq{SW-action}:
\bea
S_{SW,2} &=& 2  Tr \theta^{ab}
[X^a,A_c([Y^c,A^b] +  F^{cb})]  \nn\\ 
&=& 2 Tr \theta^{ab} \Big(
[X^a,A_c ](F^{cb} + [Y^c, A^b])  
+ A_c [X^a,(\cF^{cb} + [Y^c, A^b] 
+ A_d [X^d,\theta^{cb}])] \Big) \nn\\ 
&=& 2 Tr \theta^{ab} \Big(
[X^a,A_c ](F^{cb} + [Y^c, A^b]) 
+ A_c [X^a,(\cF^{cb} + [Y^c, A^b] 
+ A_d [X^d,\theta^{cb}])] \Big) \nn
\eea
where we used \eq{F-identity}
\be
\cF^{ab} = F^{ab} - A_d[Y^d,\theta^{ab}] \,\, + O(\theta^3),
\ee
noting \eq{AA-drop}.
The second line can be simplified using \eq{acd-id-3}
\be
\theta^{ab} [X^a,[Y^c,A^b]]
=\frac 12\theta^{ab} [X^c,\cF^{ab}] 
- \frac 12\theta^{ab} [A^c,F^{ab}] 
- \theta^{ab}[A^a,\theta^{cb}] 
\ee
and \eq{thetadF-id}
\be
\theta^{ab} [X^a,\cF^{cb}] = \frac 12 \theta^{ab} ([X^c,\cF^{ab}]+[A^c,\theta^{ab}]) 
-  \theta^{ab}[A^a,\theta^{cb}] 
\ee
so that
\bea
S_{SW,2} &=&   Tr \theta^{ab} \Big(2[X^a,A_c ](F^{cb} + [Y^c, A^b])  \nn\\ 
&&+  2A_c [X^c,\cF^{ab}]
- A_c\theta^{ab} [A^c,F^{ab}] 
+   A_c [A^c,\theta^{ab}] - 4 A_c [A^a,\theta^{cb}] \nn\\ 
&&+   2 A_c [X^a,A_d] [Y^d,\theta^{cb}] 
  +  2 A_c A_d [X^a,[X^d,\theta^{cb}]]\Big) \,.
\eea
Now
\bea
  \theta^{ab} A_c A_d [Y^a,[Y^d,\theta^{cb}]] 
&=&  \theta^{ab} A_c A_d [Y^d,[Y^a,\theta^{cb}]] 
  +  \theta^{ab} A_c A_d [\theta^{ad},\theta^{cb}] \nn\\
&=&  \frac 12 Tr \theta^{ab} A_c A_d  [Y^d,[Y^c,\theta^{ab}]]
  +  \theta^{ab} A_c A_d [\theta^{ad},\theta^{cb}]
\eea
using \eq{aayy-id},
which implies
\be
  \theta^{ab} A_c A_d [X^a,[X^d,\theta^{cb}]] 
=  \frac 12 Tr \theta^{ab} A_c A_d  [X^d,[X^c,\theta^{ab}]]
  +  \theta^{ab} A_c A_d [\theta^{ad},\theta^{cb}] \,\,+ O(\theta^5)
\ee
hence
\bea
S_{SW,2} &=&  Tr \theta^{ab} \Big(
2 [Y^a,A_c ](F^{cb} + [Y^c, A^b])  \nn\\ 
&+&  2 A_c [X^c,\cF^{ab}]
- A_c\theta^{ab} [A^c,F^{ab}] 
+   A_c [A^c,\theta^{ab}]- 4A_c [A^a,\theta^{cb}] \nn\\ 
&+&  2 A_d [X^a,A_c] [Y^c,\theta^{db}] 
 +  A_c A_d  [X^d,[X^c,\theta^{ab}]]
  +  2 A_c A_d [\theta^{ad},\theta^{cb}]
\Big)\,.
\eea
The first line can be written using \eq{FF-id-2-1} 
which gives
\bea
S_{SW,2} &=& Tr  \theta^{ab} \Big( 2 F^{ad}\theta^{-1}_{dc} F^{bc} 
+  2 F^{ad} \theta^{-1}_{dc}A_e [Y^c,\theta^{eb}] 
 +   2 \theta^{ae}  i\{A_e, A^b\}\nn\\ 
&+& 2  A_c [X^c,\cF^{ab}]
- A_c\theta^{ab} [A^c,F^{ab}] 
+  A_c [A^c,\theta^{ab}]- 4 A_c [A^a,\theta^{cb}] \nn\\ 
&+&  2 A_d [X^a,A_c] [Y^c,\theta^{db}] 
 +  A_c A_d  [X^d,[X^c,\theta^{ab}]]
  + 2 A_c A_d [\theta^{ad},\theta^{cb}] \Big)\,.
\eea
Now using \eq{FA-comb} this becomes
\bea
S_{SW,2} &=&  Tr \theta^{ab} \Big(2 F^{ad}\theta^{-1}_{dc} F^{bc} 
+  2\theta^{ae} i \{A_e, A^b\} 
 +  2\theta^{ae} A_d [A_e,\theta^{db}]  \nn\\ 
&&+ 2 A_c [X^c,\cF^{ab}]- A_c\theta^{ab} [A^c,F^{ab}] 
+  A_c [A^c,\theta^{ab}]- 4 A_c [A^a,\theta^{cb}] \nn\\ 
&&+  A_c A_d  [X^d,[X^c,\theta^{ab}]] +  2 A_c A_d [\theta^{ad},\theta^{cb}]
\Big) \,.\nn 
\eea
Replacing $Tr\theta^{ab}\theta^{ae} i \{A_e, A^b\} 
\to Tr\theta^{ab}\theta^{ae} [A_e, A^b]$ and noting
\bea
&&  \theta^{ab} \(\theta^{ae}  [A_e, A^b] 
 +  \theta^{ae} A_d [A_e,\theta^{db}] 
- 2 A_c [A^a,\theta^{cb}] 
 +  A_c A_d [\theta^{ad},\theta^{cb}] \)\nn\\ 
&&= \theta^{ab}\(\theta^{ad}  [A_d, A^b] 
 + \theta^{ad} A_c [A_d,\theta^{cb}] 
- 2A_c [A^a,\theta^{cb}]  
+  A_c [A_d\theta^{ad},\theta^{cb}]
  - \theta^{ad} A_c [A_d,\theta^{cb}]\)\nn\\ 
&&=  \theta^{ab}\theta^{ad}  [A_d, A^b] 
- \theta^{ab}A_c [A^a,\theta^{cb}] \nn\\ 
&&=  \theta^{ab}\theta^{ad}  [A_d, A^b]
- \theta^{ab} A_d [\theta^{da},A^b] \nn\\
&&=  \theta^{ab}[A^a, A^b] 
\eea
(note: only the abelian component involving the Poisson bracket
contributes)
and
\be
Tr\,\Big( A_c\theta^{ab} [A^c,F^{ab}] \Big)
= - Tr\,\Big( [A^c,A_c]\theta^{ab} F^{ab} \Big)
\ee
(since only the nonabelian terms survive), we obtain
\bea
S_{SW,2} &=&  Tr \theta^{ab} \Big(2 F^{ad}\theta^{-1}_{dc} F^{bc} 
+  2 [A^a, A^b]+  A_c [A^c,\theta^{ab}] \nn\\ 
&& + 2 A_c [X^c,\cF^{ab}] + [A^c,A_c]\theta^{ab} F^{ab}
+ A_c A_d  [X^d,[X^c,\theta^{ab}]]
\Big) \,.
\eea
Now we use 
\bea
  A_c [X^c,\cF^{ab}] &=& 
 A_c [X^c,F^{ab} - A_d [Y^d,\theta^{ab}]] \nn\\
&=&   A_c [X^c,F^{ab}] - A_c[X^c, A_d [Y^d,\theta^{ab}]]\nn\\
&=&   A_c [X^c,F^{ab}] - A_c[X^c, A_d [X^d,\theta^{ab}]]
\eea
(to $O(\theta^4)$)
using \eq{F-identity},
and obtain
\bea
S_{SW,2} &=& Tr \theta^{ab} \Big(2 F^{ad}\theta^{-1}_{dc} F^{bc} 
+ 2\theta^{ab}[A^a, A^b]  + 2 A_c [X^c,F^{ab}] 
+[A^c,A_c]\theta^{ab} F^{ab}
+  A_c [A^c,\theta^{ab}] \nn\\ 
&& +  A_c A_d  [X^d,[X^c,\theta^{ab}]] 
 - 2 A_c[X^c, A_d [X^d,\theta^{ab}]] 
\Big) .
\label{SW-3}
\eea
Using partial integration, we have
\be
 Tr \theta^{ab} A_d [X^d,A_c[X^c,\theta^{ab}]]  
= -Tr A_d [X^d,\theta^{ab}] A_c [X^c,\theta^{ab}]
 - Tr \theta^{ab} [X^d,A_d]  A_c [X^c,\theta^{ab}]  
\ee
and
\bea
Tr \theta^{ab} A_c A_d [X^d,[X^c,\theta^{ab}]]
&=& Tr \theta^{ab}  A_d [X^d,A_c[X^c,\theta^{ab}]]
   - Tr \theta^{ab}  A_d [X^d,A_c][X^c,\theta^{ab}] \nn\\
&=& Tr -\theta^{ab}  [X^d,A_d] A_c [X^c,\theta^{ab}]
  - [X^d,\theta^{ab}]  A_d A_c[X^c,\theta^{ab}] \nn\\
 && - Tr \theta^{ab}  A_d [X^d,A_c][X^c,\theta^{ab}] \nn
\eea 
therefore
\bea
&& Tr -2\theta^{ab} A_d [X^d,A_c[X^c,\theta^{ab}]]  
+  \theta^{ab} A_c A_d [X^d,[X^c,\theta^{ab}]]   \nn\\
&=& Tr A_d [X^d,\theta^{ab}] A_c [X^c,\theta^{ab}]
 + \theta^{ab} [X^d,A_d]  A_c [X^c,\theta^{ab}] 
 - \theta^{ab}  A_d [X^d,A_c][X^c,\theta^{ab}] \,.\nn
\eea
Consider the term
\bea
-  2Tr \theta^{ab}  A_d [X^d,A_c][X^c,\theta^{ab}] 
 &=& Tr \theta^{ab}  [X^c,A_d] [X^d,A_c]\theta^{ab}
 + \theta^{ab} \theta^{ab} A_d [X^c,[X^d,A_c]]\nn\\
 &=& Tr \theta^{ab}  [X^c,A_d] [X^d,A_c]\theta^{ab}
 + \theta^{ab} \theta^{ab} A_d [X^d,[X^c,A_c]] \nn\\
&& + \theta^{ab} \theta^{ab} A_d [(\theta^{cd} + \cF^{cd}),A_c]\nn\\
 &=& Tr \theta^{ab}  [X^c,A_d] [X^d,A_c]\theta^{ab}
 - \theta^{ab} \theta^{ab} [X^d,A_d] [X^c,A_c] \nn\\
 && -2 \theta^{ab} A_d [X^d,\theta^{ab}]  [X^c,A_c] 
 + \theta^{ab} \theta^{ab} A_d [(\theta^{cd} + \cF^{cd}),A_c] \nn
\eea
(using partial integration again),
which gives
\bea
&& - Tr \theta^{ab}  A_d [X^d,A_c][X^c,\theta^{ab}] 
+ \theta^{ab} A_d [X^d,\theta^{ab}]  [X^c,A_c]  \nn\\
&&= Tr \frac 12 \theta^{ab}  [X^c,A_d] [X^d,A_c]\theta^{ab}
 - \frac 12\theta^{ab} \theta^{ab} [X^d,A_d] [X^c,A_c] 
 + \frac 12\theta^{ab} \theta^{ab} A_d [(\theta^{cd} + \cF^{cd}),A_c] \nn
\eea
and we obtain
\bea
&& Tr -2\theta^{ab} A_d [X^d,A_c[X^c,\theta^{ab}]]  
+  \theta^{ab} A_c A_d [X^d,[X^c,\theta^{ab}]]   \nn\\
&=& Tr A_d [X^d,\theta^{ab}] A_c [X^c,\theta^{ab}]
+ \frac 12 \theta^{ab}  [X^c,A_d] [X^d,A_c]\theta^{ab}
 - \frac 12\theta^{ab} \theta^{ab} [X^d,A_d] [X^c,A_c] \nn\\
&& + \frac 12\theta^{ab} \theta^{ab} A_d [(\theta^{cd} + \cF^{cd}),A_c] \,.
\eea
Inserting this into \eq{SW-3} and using
\bea
&& Tr\(\theta^{ab}  A_c [A^c,\theta^{ab}] 
-\frac 12\theta^{ab}\theta^{ab}  A_c [\theta^{cd},A_d] \) \nn\\
&&= Tr\(\theta^{ab}  A_d [A^d,\theta^{ab}] 
+ \frac 12\theta^{ab}\theta^{ab} [A^d,A_d] 
+ \frac 12\theta^{ab}\theta^{ab} \theta^{cd} [A_c ,A_d]\)  \nn\\
&&= Tr\(\frac 12\theta^{ab}\theta^{ab} \theta^{cd} [A_c ,A_d]\)
\eea
(again only the abelian contribution from the Poisson-bracket survives)
gives
\bea
S_{SW,2} &=&    Tr  \Big(2 \theta^{ab}F^{ad}\theta^{-1}_{dc} F^{bc} 
+ 2\theta^{ab}[A^a, A^b]
 + 2\theta^{ab}  A_c [X^c,F^{ab}] 
+[A^c,A_c]\theta^{ab} F^{ab} \nn\\
&& + \frac 12\theta^{ab} \theta^{ab}A_d [ \cF^{cd},A_c]
 +  A_d [X^d,\theta^{ab}] A_c [X^c,\theta^{ab}] \nn\\
&& + \frac 12\theta^{ab} \theta^{ab} 
([X^d,A_c] [X^c,A_d]  -[X^d,A_d] [X^c,A_c] 
+ \theta^{cd} [A_c ,A_d] ) \Big) \,. \nn\\
\eea
Now observe that 
\be
[Y^c,A_d] [Y^d,A_c] - \theta^{cd}[A_d,A_c] 
= \theta^{ci} \theta^{dj} \partial_i A_d \partial_j A_c 
 - \theta^{cd} \theta^{ij} \partial_i A_d \partial_j A_c 
- \theta^{cd} [A_d,A_c]_{n.a.}
\ee
where $[A_d,A_c]_{n.a.}$ stands for commutator of the nonabelian
components. We can drop terms of the type $Tr \theta^4\, f(x) [A_a,A_b]_{n.a.}$
under the trace. 
Therefore 
\bea
[X^c,A_d] [X^d,A_c] - \theta^{cd}[A_d,A_c] 
&=& \theta^{ci} \theta^{dj} (\partial_i A_d + [A_i, A_d])
   (\partial_j A_c + [A_j, A_c])
 - \theta^{cd} \theta^{ij} \partial_i A_d \partial_j A_c  \nn\\
&=& \theta^{ci} \theta^{dj} \Big(
\partial_i A_d\partial_j A_c + \partial_i A_d [A_j, A_c] \nn\\
&& + [A_i, A_d] \partial_j A_c + [A_i, A_d] [A_j, A_c]
 -  \partial_d A_i \partial_j A_c \Big) \nn\\
%&=& \theta^{ci} \theta^{dj} \Big(
%(\partial_i A_d -  \partial_d A_i)\partial_j A_c 
%+ 2 [A_i, A_d] \partial_j A_c + [A_i, A_d] [A_j, A_c] \Big) \nn\\
% &=& \theta^{ci} \theta^{dj} \Big(
% \frac 12 (\partial_i A_d -  \partial_d A_i)
% (\partial_j A_c -  \partial_c A_j) \nn\\
% && +  [A_i, A_d] (\partial_j A_c - \partial_c A_j) 
%  + [A_i, A_d] [A_j, A_c] \Big) \nn\\
&=& - \frac 12 \theta^{cd} \theta^{ij} F_{id} F_{jc} 
  - \frac 12 \theta^{cd} \theta^{ij} [A_i, A_d] [A_j, A_c]
\eea
since
$\theta^{ci} \theta^{dj} [A_j, A_c] \partial_i A_d 
=\theta^{ci} \theta^{dj} [A_i, A_d] \partial_j A_c$ under the trace.
Furthermore,
\be
[Y^a,A_a] = \theta^{ab} \partial_b A_a 
= -\frac 12 \theta^{ab} (\partial_a A_b  -\partial_b A_a )
\ee
and therefore 
\bea
[X^a,A_a] &=& -\frac 12 \theta^{ab} (\partial_a A_b  -\partial_b A_a)
 + \theta^{ab} [A_b,A_a] \nn\\
&=& - \frac 12 \theta^{ab} (F_{ab} + [A_a,A_b])
\eea
or 
\bea
2 \theta^{ab} [X^c,A_c] F^{ab} - [A^c,A_c]\theta^{ab} F^{ab}
= - \theta^{cd} F_{cd} \theta^{ab}F^{ab}
\label{FF-id-2}
\eea
hence 
\bea
S_{SW,2}
 &=& Tr \Big(2 \theta^{ab}F^{ad}\theta^{-1}_{dc} F^{bc} 
+ 2\theta^{ab}[A^a, A^b]
 + 2\theta^{ab}  A_c [X^c,F^{ab}] 
+[A^c,A_c]\theta^{ab} F^{ab} \nn\\
&& + \frac 12\theta^{ab} \theta^{ab}A_d [ \cF^{cd},A_c]
 +  A_d [X^d,\theta^{ab}] A_c [X^c,\theta^{ab}]  \nn\\ 
&&  - \frac 18\theta^{ab}\theta^{ab}\theta^{cd}\theta^{ij}
\Big(F_{cd}F_{ij} + 2 F_{id} F_{jc} + 2 F_{cd}[A_i, A_j] \Big) \Big) 
\eea
where we used
\bea
&& Tr \theta^{ab}\theta^{ab}\theta^{cd} \theta^{ij}\Big(2  [A_i, A_d] [A_j, A_c]
 + [A_i, A_j] [A_c, A_d] \Big) \nn\\
&& = -Tr \theta^{ab}\theta^{ab}\theta^{cd} \theta^{ij}\Big(2 A_i [[A_j, A_c], A_d] 
 + A_i [[A_c, A_d], A_j]  \Big) \nn\\
&& =  -Tr \theta^{ab}\theta^{ab}\theta^{cd} \theta^{ij}\Big(2 A_i [[A_j, A_c], A_d] 
 -  A_i [[A_d, A_j], A_c]  - A_i [[A_j, A_c], A_d] \Big) =0 \,.\nn
%&& =  -Tr \theta^{ab}\theta^{ab}\theta^{cd} \theta^{ij}\Big(2 A_i [[A_j, A_c], A_d] 
% + A_i [[A_c, A_j], A_d]  - A_i [[A_j, A_c], A_d] \Big) \nn\\
%&& = 0
\eea
Together with \eq{action-expanded}, we obtain
\bea
S &=& -Tr \Big(F^{ab}F^{ab} - 2 F^{ab} A_c [Y^c,\theta^{ab}]
+ A_c [Y^c,\theta^{ab}] [Y^d,\theta^{ab}] A_d
 + 2 \theta^{ab}[A^a,A^b]\Big) + S_{SW,2}  \nn\\
&=& -Tr \Big(F^{ab}F^{ab} - 2 F^{ab} A_c [X^c,\theta^{ab}]
+ A_c [X^c,\theta^{ab}] [X^d,\theta^{ab}] A_d
 + 2 \theta^{ab}[A^a,A^b] \Big) + S_{SW,2}\,. \nn
\eea
Replacing $Y \to X$ which is correct to $O(\theta^4)$,
we obtain 
\bea
S  &=& -Tr \Big(F^{ab} F^{ab} 
+ 2  A_c [X^c,F^{ab}] \theta^{ab}
+ 2  \theta^{ab} [X^c,A_c] F^{ab}
 + [X^c,\theta^{ab}] A_c [X^d,\theta^{ab}] A_d  \nn\\
&& + 2 Tr \theta^{ab} [A^a,A^b] \Big) + S_{SW,2} \nn\\
%  &=& -Tr \Big(F^{ab} F^{ab} 
% + 2  A_c [X^c,F^{ab}] \theta^{ab}
% + 2  \theta^{ab} [X^c,A_c] F^{ab}
%  + [X^c,\theta^{ab}] A_c [X^d,\theta^{ab}] A_d  \nn\\
% && + 2 \theta^{ab} [A^a,A^b] 
% - \Big(2 \theta^{ab}F^{ad}\theta^{-1}_{dc} F^{bc} 
% + 2\theta^{ab}[A^a, A^b]
%  + 2\theta^{ab}  A_c [X^c,F^{ab}] 
% +[A^c,A_c]\theta^{ab} F^{ab} \nn\\
% && + \frac 12\theta^{ab} \theta^{ab}A_d [\cF^{cd},A_c]
%  +  A_d [X^d,\theta^{ab}] A_c [X^c,\theta^{ab}]  \nn\\ 
% &&  - \frac 18\theta^{ab}\theta^{ab}\theta^{cd}\theta^{ij}
% \(F_{cd}F_{ij} + 2 F_{id} F_{jc} + 2 F_{cd}[A_i, A_j] \) \Big)\Big) \nn\\
&=& -Tr \Big(F^{ab} F^{ab} 
 - \theta^{ab}  F^{ab}\theta^{cd} F_{cd}
-  2 \theta^{ab}F^{ad}\theta^{-1}_{dc} F^{bc}  \nn\\
&& + \frac 18\theta^{ab}\theta^{ab}\theta^{cd}\theta^{ij}
\Big(F_{cd}F_{ij} + 2 F_{id} F_{jc} + 2 F_{cd}[A_i, A_j] \Big) 
- \frac 12\theta^{ab} \theta^{ab} A_d [ \cF^{cd},A_c] \Big) \,.
\eea
Finally we use \eq{AFA-id-2} together with \eq{FF-id-2} which gives
\be
Tr \theta^{ab} \theta^{ab} A_d [\cF^{cd},A_c]
= \frac 12 Tr \theta^{ab} \theta^{ab}\theta^{ij}[A_i, A_j] \theta^{cd} F_{cd}
\ee
and we obtain the gauge invariant action
\bea
S &=& -Tr \Big(F^{ab} F^{ab} 
 - \theta^{ab}  F^{ab}\theta^{cd} F_{cd}
-  2 \theta^{ab}F^{ad}\theta^{-1}_{dc} F^{bc} 
 + \frac 18\theta^{ab}\theta^{ab}\theta^{cd}\theta^{ij}
\big(F_{cd}F_{ij} + 2 F_{id} F_{jc} \big)  \Big) \nn\\
&=&  -Tr F^{ab} F^{ab} + S_{NC}  
\label{action-expanded-2-app}
\eea
Needless to say that there should be a simpler way to obtain this.

\section{Appendix B: Newtonian metric}

We want to reproduce the metric \eq{newton-metric} in terms of $h_{ij}$
\eq{flat-grav-wave-3}.
$F_{ab}$ is a function of  $(y^0,y^1,y^2,y^3)$
with $\eta_{ab} = (-1,1,1,1)$ and has the form
\be
F_{ab} = \left(\begin{array}{cccc} 0 & E_1 & E_2 & E_3  \\
                                -E_1 & 0 & B_3 & -B_2 \\
                                -E_2 & -B_3 & 0 & B_1  \\
                                -E_3 & B_2 & -B_1 & 0
                              \end{array}\right) \,.
\ee
We can assume that
$\theta^{ab} = \theta\left(\begin{array}{cccc} 0 & 0 & 0 & 1 \\
                                0 & 0 & 1 & 0 \\
                                0 & -1 & 0 & 0  \\
                                -1 & 0 & 0 & 0 \end{array}\right)$,
which gives
\be
h_{ab} = \theta^{-1}\left(\begin{array}{cccc} 2 E_3 & -B_2-E_2 & B_1+E_1 & 0 \\
                                 -B_2-E_2 & -2B_3 & 0 & B_1-E_1 \\
                                B_1+E_1 & 0 & -2B_3 & B_2-E_2  \\
                                0 & B_1-E_1 & B_2-E_2 & -2E_3 \end{array}\right)
\ee
and $\obar g_{ab} = \theta^{-2}(-1,-1,-1,1)$, so that $y^3$ turns into the 
time $t$.
Since we want the metric to be static i.e. time-independent and invariant
under time reflections, we  consider an electromagnetic field which is independent of
$y^3$, $\partial_3 F_{ab} =0$, and require
\be
B_1 = E_1, \quad B_2 = E_2 .
\label{BE-constraint}
\ee
Then
\be
h_{ab} = 2\theta^{-1}\left(\begin{array}{cccc} E_3 & -E_2 & E_1 & 0 \\
                                 -E_2 & -B_3 & 0 & 0 \\
                                E_1 & 0 & -B_3 & 0 \\
                                0 & 0 &0 & -E_3 \end{array}\right)
\ee
where as usual $E_i$ and $B_i$ 
can be written in the form
\be
E_i = \partial_0 A_i - \partial_i A_0, \qquad
B_i = \varepsilon_{ijk}\, \partial_j A_k 
\ee
and the derivatives are w.r.t. $y^a$.
The Bianci identities are
\be
\partial_i B_i=0, \qquad 
\varepsilon_{ijk}\, \partial_j E_k - \partial_0 B_i =0 \, .
\ee
Since we want to consider static configurations we have
$\partial_3 B_3 =0$ (recall $t=y^3$), hence
\be
\partial_1
B_1 + \partial_2 B_2=0 .
\label{B-constraint}
\ee
Now  fix the gauge by setting $A_3=0$ (cf. axial gauge). Then 
\be
B_1 = -\partial_3 A_2, \quad 
B_2 = \partial_3 A_1, \quad E_3 =  -\partial_3 A_0 \, ,
\ee
which can be solved for arbitrary $B_1,B_2,E_3$ satisfying the Bianci identities by
\bea
A_2 &=& -y_3 B_1(y^0,y^1,y^2) + \tilde A_2(y^0,y^1,y^2), \nn\\
A_1&=& y_3 B_2(y^0,y^1,y^2)+ \tilde A_1(y^0,y^1,y^2), \nn\\
A_0 &=& -y_3 E_3(y^0,y^1,y^2) + \tilde A_0(y^0,y^1,y^2) 
\eea
with arbitrary $\tilde A_{0,1,2}(y^0,y^1,y^2)$.
Then $E_1, E_2$ can be computed as
\bea
E_1 &=& -\partial_1 \tilde A_0 + \partial_0 \tilde A_1, \nn\\
E_2 &=& -\partial_2 \tilde A_0 + \partial_0 \tilde A_2, 
\eea
where the $y^3$-dependent terms vanish due to the Bianci identity.

The most general $B_i$ satisfying \eq{B-constraint} can be written as
\be
B_1 = \partial_2 \phi,\quad B_2 = -\partial_1 \phi
\label{B-U-relation}
\ee
for any given $\phi(y^0,y^1,y^2)$. 
Setting $\phi = \partial_0 \varphi$ and defining
$\tilde A_0=0,\, \tilde A_1 = \partial_2\varphi,\, \tilde A_2 = -\partial_1
\varphi$ 
we get indeed $E_1=B_1,\, E_2=B_2$ and 
\be
B_3= \partial_1 \tilde A_2 - \partial_2 \tilde A_1 = - \Delta_{12}\varphi.
\ee
$E_3$ is almost determined by the Bianci identity, which is solved by
\be
E_3=\partial_0 \phi = \partial_0^2 \varphi \,.
\ee
Now perform a change of variables ${y^a}' = y^a + \theta\xi^a$
with $\xi^a = 2(\phi,0,0,0)$,
which gives
\be
h_{ab}' =  2\theta^{-1}\left(\begin{array}{cccc} - E_3 & 0& 0& 0 \\
                                0 & -B_3 & 0 & 0 \\
                                0 & 0 & -B_3 & 0 \\
                                0 & 0 &0 & -E_3 \end{array}\right) \, .
\label{h-newton}
\ee
Assuming that $O(B_3) \approx O(\partial_0 \phi)$, this describes Newtonian
gravity with gravitational potential given by 
\be
U(y^0,y^1,y^2) = \theta E_3 = \theta \partial_0^2 \varphi
\label{U-def}
\ee
which is arbitrary since $U$ is arbitrary.
It can therefore describe an arbitrary static mass distribution $\rho$ by
solving the Poisson equation
\be
\Delta_{(3)} U  = 4\pi G \rho,
\ee
which is expected to follow from the gravity action.
For the vacuum $\rho =0$, and $E_3 = B_3$ follows from 
$\Delta_{(3)} \varphi=0$ (up to a constant), in agreement with general
relativity.

\section{Appendix C}

One way to show \eq{eta-eval} is to note that
\bea
(\tilde\theta\wedge\theta)^{ijkl} 
&=&  (\tilde\theta^{ij} \theta^{kl} 
- \tilde \theta^{il}\theta^{kj} - \tilde\theta^{lj} \theta^{ki})
+ (\tilde\theta^{kl}\theta^{ij} 
- \tilde\theta^{kj}\theta^{il}  - \tilde\theta^{ki}\theta^{lj})
\eea
and to consider
\bea
(\theta^{-1}\wedge \theta^{-1})_{ijkl} (\tilde\theta\wedge\theta)^{ijkl} &=& 
(\theta^{-1}\wedge \theta^{-1})_{ijkl}\tilde\theta^{ij} \theta^{kl} \nn\\
&=&  (\theta^{-1}_{ij} \theta^{-1}_{kl} - \theta^{-1}_{il} \theta^{-1}_{kj} 
- \theta^{-1}_{lj} \theta^{-1}_{ki}) \tilde\theta^{ij} \theta^{kl} \nn\\
&=& (\theta^{-1}_{ij}\tilde\theta^{ij}) (\theta^{-1}_{kl}\theta^{kl})
 +  2 (\theta^{-1}_{il} \theta^{-1}_{jk} \tilde\theta^{ij}\theta^{kl})\nn\\
&=&  (d - 2) \theta^{js}\theta^{sj}
 =  (d - 2) G^{ab} g_{ab}  
\eea
where $d=4$ is the dimension of space(time). On the other hand,
\be
(\theta^{-1}\wedge \theta^{-1})_{ijkl} (\theta\wedge\theta)^{ijkl}
= (\theta^{-1}_{ij}\theta^{ij}) (\theta^{-1}_{kl}\theta^{kl})
 +  2 (\theta^{-1}_{il}\theta^{-1}_{jk}\theta^{ij}\theta^{kl})\nn\\
=  d( d- 2)
\ee
which together implies \eq{eta-eval}.

\end{document}